\documentclass{amsart}

\usepackage{amsmath}
\usepackage{amsthm}
\usepackage{amsfonts}
\usepackage{amssymb}
\usepackage{graphicx}
\usepackage{xcolor}
\usepackage{setspace}
\usepackage{subcaption}
\usepackage{etoolbox}
\usepackage[foot]{amsaddr}
\usepackage{etoolbox}
\usepackage{tikz}
\usepackage{mathtools} 
    
\usepackage{multirow}
\usepackage{multicol}
\usepackage{array}
\usepackage{nicefrac}       

\usepackage{stmaryrd}

\newtheorem{theorem}{Theorem}[section]

\newtheorem{lemma}[theorem]{Lemma}
\newtheorem{proposition}[theorem]{Proposition }

\usepackage[authoryear]{natbib}

\usepackage{hyperref}
\hypersetup{
    colorlinks,
    linkcolor={red!50!black},
    citecolor={blue!50!black},
    urlcolor={blue!80!black}
}

\makeatletter
    \patchcmd{\NAT@test}{\else \NAT@nm}{\else \NAT@nmfmt{\NAT@nm}}{}{}
    \DeclareRobustCommand\citepos
        {\begingroup
        \let\NAT@nmfmt\NAT@posfmt
        \NAT@swafalse\let\NAT@ctype\z@\NAT@partrue
        \@ifstar{\NAT@fulltrue\NAT@citetp}{\NAT@fullfalse\NAT@citetp}}
    \let\NAT@orig@nmfmt\NAT@nmfmt
    \def\NAT@posfmt#1{\NAT@orig@nmfmt{#1's}}
\makeatother

\title[The Tragedy of the AI Commons]{The Tragedy of the AI Commons}

\author{Travis LaCroix$^{1,2}$ \\ Aydin Mohseni$^{3}$}

\address{\tiny{$^1$Mila - Qu{\'e}bec AI Institute \& Dept. of Computer Science and Operations Research \\ Universit{\'e} de Montr{\'e}al}}
\address{\tiny{$^2$Department of Philosophy \& Schwartz Reisman Institute for Technology and Society \\ University of Toronto}}
\address{\tiny{$^3$Department of Logic and Philosophy of Science \\  University of California, Irvine}}

\email{\scriptsize{lacroixt@mila.quebec \textrm{(Corresponding);} amohseni@uci.edu}}

\date{\scriptsize{Draft of \today}}

\begin{document}

\maketitle

\begin{abstract}
    \singlespacing
    Policy and guideline proposals for ethical artificial-intelligence research have proliferated in recent years. These are supposed to guide the socially-responsible development of AI for the common good. However, there typically exist incentives for non-cooperation (i.e., non-adherence to such policies and guidelines); and, these proposals often lack effective mechanisms to enforce their own normative claims. The situation just described constitutes a social dilemma---namely, a situation where no one has an individual incentive to cooperate, though mutual cooperation would lead to the best outcome for all involved. In this paper, we use stochastic evolutionary game dynamics to model this social dilemma in the context of the ethical development of artificial intelligence. This formalism allows us to isolate variables that may be intervened upon, thus providing actionable suggestions for increased cooperation amongst numerous stakeholders in AI. Our results show how stochastic effects can help make cooperation viable in such a scenario. They suggest that coordination for a common good should be attempted in smaller groups in which the cost for cooperation is low, and the perceived risk of failure is high. This provides insight into the conditions under which we should expect such ethics proposals to be successful with regard to their scope, scale, and content.
\end{abstract}

\section{Introduction}

Artificial intelligence promises to fundamentally change nearly every facet of our lives, for better or worse \citep{Harari-2017, Helbing-2019, Makridakis-2017}. In response to this reality, there has been a proliferation of policy and guideline proposals for ethical artificial-intelligence and machine-learning (AI/ML) research. Jobin et al. \cite{Jobin-et-al-2019} survey several global initiatives for AI/ML and find no fewer than 84 documents containing ethics principles for AI research, with 88\% of these having been released since 2016. More broadly speaking, the World Economic Forum has identified almost three hundred separate efforts to develop ethical principles for AI \citep{Russell-2019}. Policy documents of this sort are meant to specify `best practices' to which engineers, developers, researchers, etc. ought to adhere. 

For example, the {\it Montr{\'e}al Declaration for the Responsible Development of AI} identifies a set of abstract normative principles and values intended to promote the fundamental interests of stakeholders; signatories are invited to commit to `the development of AI at the service of the individual and the common good' \citep{Montreal-Declaration-2017}. Proposals of this sort typically highlight issues concerning transparency, justice and fairness, non-maleficence, responsibility, and privacy, among others \citep{Jobin-et-al-2019}. These initiatives generally take one of two approaches to foster the ethical practice of AI research: proposing principles to guide the socially-responsible development of AI or examining the societal impacts of AI \citep{Luccioni-Bengio-2019}. 

However, policies-and-procedures documents, like the {\it Montr{\'e}al Declaration}, are examples of `non-legislative policy instruments' or `soft law' \citep{Sossin-Smith-2003}, which are instruments for cooperation that are not legally binding. This stands in contrast to `hard law', which consists in legally-binding regulations passed by legislatures. By definition, then, these reports are not intended to produce enforceable rules but are meant merely as {\it guides} for ethical practice. Therefore, the {\it proliferation} of such guiding principles raises pressing questions about their efficacy. It has been suggested that such declarations have little to no practical effect \citep{Hagendorf-2019}; at least one study finds that the effectiveness of ethical codes for influencing the behaviour of professionals in the technology community is virtually nonexistent \citep{McNamara-et-al-2018}. 

Part of the problem is that, in the AI research community, as in many other social contexts, there are typically 
costs associated with cooperation---e.g., the additional time and financial resources that are required to ensure research and business practices adhere to the proposed normative standards---in addition to incentives for non-cooperation or non-adherence to ethics agreements---e.g., a lucrative defence contract to develop autonomous weapons. Furthermore, non-legislative policy instruments, like guiding principles, lack effective mechanisms to {\it enforce} (or {\it reinforce}) 
their own normative claims, because they depend on voluntary, non-binding cooperation between the relevant parties---e.g., individual researchers, labs, academic bodies, corporations, governments, international bodies, etc.

This failure should come as no surprise to game theorists and economists: in the situation just described, no one has an individual incentive to cooperate, although mutual cooperation would lead to the best outcome for all those involved. This constitutes a {\it social dilemma} \citep{Allison-Kerr-1994, Dawes-1980, Hobbes-1651, Hume-1739, Ross-SEP-game-theory, Serrano-Feldman-2013}.

In this paper, we use stochastic evolutionary game dynamics to model this social dilemma in the context of the ethical development of AI. This model allows us to isolate variables that may be intervened upon; this, in turn, helps to formulate actionable suggestions for increased cooperation amongst numerous stakeholders in AI. Our results show how stochastic effects can help make cooperation viable in such a scenario, and they suggest, among other things, that coordination for a common good should be attempted in smaller groups in which the cost for cooperation is low, and the perceived risk of failure is high. This analysis provides insight into the conditions under which we may expect such policy and guideline proposals for safe and ethical AI to be successful in terms of their scope, scale, and content. This, in turn, yields plausible solutions to the social dilemmas that the AI/ML community face for acting in a socially responsible and ethical way that is aligned with a common good.\footnote{Of course, it is nontrivial to determine precisely what a `common good' is; see discussion in \cite{Green-2019}.}

In Section~\ref{sec:Backgrounder}, we discuss related work. In Section~\ref{sec:Model}, we present an evolutionary game which models cooperation (with respect to proposed ethical norms and principles in AI) using a {\it social dynamics} framework. Select results are presented in Section~\ref{sec:Results}. Section~\ref{sec:Conclusion} concludes with a number of morals we derive from our model. Here, we outline a concrete situation under which these morals may be applied, and discuss the practical applications that the insights from our model proffer.


\section{Background and Related Work}
\label{sec:Backgrounder}

In this section, we discuss related work and highlight how our model-based analysis of adherence to AI ethics guidelines, when understood as a social dilemma, provides substantive and actionable suggestions for fostering cooperation. 

\subsection*{Artificial Intelligence and Normative Guides.}

As mentioned, non-legislative policy instruments, 
in the form of ethics guidelines, have proliferated in recent years. These guidelines, codes, and principles for the responsible creation and use of new technologies come from a wide array of sources, including academia, professional associations, and non-profit organisations;\footnote{See, for example \citet{FLI-2017, Gotterbarn-et-al-2018,  HAIP-2018, ITI-2017, PAI-2016, RSS-Guideline-2019, Stanford-2018, Future-Society-2017, IEEE-2017, JSAI-2017, Public-Voice-2018, UNI-2017, Montreal-Declaration-2017, USACM-2017}.} governments;\footnote{e.g., \citet{EGE-2018, MIC-2017, MIC-2018, House-of-Lords-2018}.} and industry, including for-profit corporations.\footnote{e.g., \citet{DeepMind-2017, Google-2018, IBM-2017, IBM-2018, Microsoft-2018, OpenAI-2018, Sage-2017, SAP-2018, Sony-2018}.} Several researchers have noted that the very fact that a diverse set of stakeholders would exert such an effort to issue AI principles and policies is strongly indicative that these stakeholders have a vested interest in shaping policies on AI ethics to fit their own priorities \citep{Benkler-2019, Greene-et-al-2019, Jobin-et-al-2019, Wagner-2018}.

Perhaps unsurprisingly, it has also been widely noted that ethics policies for AI research are typically or relatively ineffective: they do not have an actual impact on human decision-making in the field of AI/ML \citep{Hagendorf-2019}; they are often too broad or high-level to guide ethics in practice \citep{Whittlestone-et-al-2019}; and they are, by their very nature, voluntary and nonbinding \citep{Campolo-et-al-AI-Now-2017, Whittaker-et-al-AI-Now-2018}.

The content of ethics guidelines can be varied to promote adherence. For example, substantial changes in abstraction can help the application, impact, and influence of ethics principles for AI by way of {\it targeted} suggestions that can be implemented easily and concretely in practice \citep{Gebru-et-al-2020, Morley-et-al-2019}. However, while several authors have noted the inefficacy of guiding principles, and have offered proposals to surmount this inefficacy, such proposals are, themselves, often couched in inherently {\it normative} language. Therefore, they too fail to instantiate any mechanisms for their own reinforcement. 

For example, ethics policies and guidelines can be understood {\it deontically}, insofar as they provide static, universal principles to be adhered to \citep{Ananny-2016, Mittelstadt-2019}. \citet{Hagendorf-2019} takes the inefficacy of such policies to be a failure of deontic ethics, proposing that AI ethics should instead be couched in a virtue-ethics framework.\footnote{Virtue ethics is a moral theory which emphasises the role of an individual's character and virtues in evaluating the rightness of actions \citep{Anscombe-1958, Aristotle-Ethics, Crisp-Slote-1997, Foot-1978}.} But these types of solutions are themselves ineffective insofar as they are merely {\it meta-level normative principles} which recommend adherence to object-level principles. Some reflection shows that we can then ask the same question of the meta-level principles as we did of the object-level ones: Should we expect {\it these} to be effective? By dint of what? Since there is no mechanism by which these meta-level principles can be enforced, they provide no solution to the inefficacy created by the proliferation of principles and guides for ethical AI. Instead, the problem is merely reproduced a level up.

In asking how the application and fulfilment of AI ethics guidelines can be improved, we shift the focus from the {\it content} of the principles and related meta-ethical considerations to the {\it cooperative}, {\it social aspects} of the dilemma. Namely, we examine the circumstances under which we should expect cooperative success to be possible or likely in a socio-dynamical context. 
This analysis has significant practical implications for regulatory principles in AI/ML, which are made all the more pressing as these technologies are increasingly integrated into society.



\section{A Socio-dynamic Model of Cooperation for Ethical AI Research}
\label{sec:Model}

In this section, we describe the model that we employ using standard techniques from evolutionary game theory. (See Appendix~\ref{App:GT} for some additional background details.) This formal framework has been applied to social dilemmas like climate change \citep{Finus-2008, Pacheco-et-al-2014, Wagner-2001} nuclear proliferation \citep{Brams-Kilgour-1987-a, Brams-Kilgour-1987-b, Kraig-1999, Zagare-1987}, conflicts over water resources \citep{Madani-2010}, etc. However, it has yet to be applied to the unique cooperative challenges that are faced in the context of AI safety and ethics.\footnote{While some of the insights from these other social dilemmas may apply here, it is hardly obvious that these situations are similar enough to the AI context to warrant wholesale transfer of the conclusions of those models to the particular situation we discuss here.}  We begin by discussing the mathematical structure of the model, including the payoff structure and dynamics. In Section~\ref{sec:Results}, we investigate the effect of several variables on the possibility or likelihood of cooperation in the context of non-legislative policy instruments for AI research. 

\subsection*{Model and Parameters.}

A population of size $Z$ is organised (divided) into groups of size $N$. Each individual in the population may be interpreted as an AI researcher who can realise or ignore some proposed norm(s) of research and practice. The population can be thought of as the global AI community at large, and each group of size $N$ in the population can be thought of as an organisational unit---e.g., a research laboratory, a university department, a collaborative research group, etc.---which can sustain shared norms. Individuals are classified according to the strategy that they choose with respect to the proposed norms for ethical behaviour within the community: they can choose either to cooperate ($C$), by adhering to the proposed norm, or to defect ($D$), by flouting the proposed norm.\footnote{We assume here that individuals who choose to cooperate, or who say they will cooperate, in fact do so.} 

Each individual has an initial endowment, $b \in \mathbb{R}^+$. Contributing to a collective, cooperative effort imposes a cost, $c \in [0,1]$, to the cooperator, consisting in a fraction of their endowment; defecting requires no such contribution. The cost may be construed as an additional effort relative to the non-cooperative baseline---e.g., taking extra precautions to ensure research and business practices adhere to the proposed normative standards---or in terms of opportunity costs---e.g., by refusing lucrative projects that would violate the proposed normative standards.

A normative agreement is successful if the fraction of contributors exceeds some threshold, $p^{*} \in (0,1)$.\footnote{A social threshold is common to many public endeavours---for example, international agreements that require a ratification threshold to take effect \citep{Chen-et-al-2012, Gokhale-Traulsen-2010, Pacheco-et-al-2009, Pacheco-et-al-2014, Santos-Pacheco-2011, Souza-et-al-2009, Wang-et-al-2009}.} If this cooperative threshold is not achieved, then with some probability, $r$, everyone in the group loses a fraction, $m$, of their endowment. $r$ can be interpreted as the {\it perceived} risk of negative consequences in the event of a failure to cooperate, and $m$ can be understood as the {\it perceived} magnitude of those consequences. That is, when $m=1$ (the failure to cooperate is perceived to be maximally impactful), individuals risk losing all of their endowment; $m=0$ represents the situation in which failure to reach an agreement has no perceived negative consequence.\footnote{Note that perceived risk of collective failure has proved important for successful collective action in dilemmas of this sort \citep{Milinski-et-al-2008, Pacheco-et-al-2014, Santos-Pacheco-2011}.}

This provides a five-dimensional space to analyse the possibility of cooperation in our social dilemma: the size of the groups in which cooperation is attempted, $N$; the perceived risk of negative consequences when cooperation fails, $r$; the perceived magnitude of said consequences, $m$; the cost of cooperation, $c$; and the critical threshold of cooperators required to avoid negative consequences, $p^{*}$. 

\subsection*{Payoffs.}

The payoffs for each strategy determine the game. An individual's payoff depends upon what action the individual chooses and what everyone else in the community is doing. 
The payoffs to cooperators in a group of size $N$ when $n_C$ individuals are cooperating is given in Equation~\ref{eq:Payoff-C},
\begin{align}\label{eq:Payoff-C}
    \pi_{C} (n_C) = b \cdot \Theta (k) + b \cdot (1 - rm) \cdot (1-\Theta (k)) - cb, \qquad 
                k = n_{C} - n^{*}.
\end{align}
$n^{*}=\lceil p^{*} \cdot N \rceil$ is the critical number of cooperators in a group, and $\Theta$ is the Heaviside step function.\footnote{That is, $\Theta(k) = 1$ when $k \geq 0$, and $\Theta(k) = 0$ otherwise.}

Equation~\ref{eq:Payoff-C} has two key terms. The left-hand summand ($b \cdot \Theta(k)$) provides payoff $b$ to the cooperator just in case $k = n_{C} - n^{*} \geq 0$; this is the payoff when cooperation succeeds. When cooperation fails ($n_{C} - n^* \leq 0$), the right-hand summand ($b \cdot (1 - rm) \cdot \Theta (k)$) comes into play. This payoff is weighted by the cost of failure as a function of the risk and magnitude of some negative consequence, $(1 - rm)$. So, when $r$ and $m$ are maximal ($rm = 1$), the entire endowment is lost if cooperation fails; when the risk or magnitude are nonexistent ($rm = 0$), none of the endowment is lost. Finally, the cost of cooperation, $cb$, is subtracted from the payoff for successful cooperation 
or failure to cooperate. 
The payoff to defectors is defined in terms of the payoff to cooperators, as in Equation~\ref{eq:Payoff-D}:
\begin{equation}\label{eq:Payoff-D}
    \pi_{D} (n_C) = \pi_C (n_C) + cb.
\end{equation}

We calculate the average payoff to each type, $C, D$, as a function of the group size, $N$, and the fraction of each type in the broader population, $x_{C}$ and $x_{D} = 1 - x_{C}$, respectively. First, we find the fraction of cooperators in the population, $x_{C}=n_C^Z/Z$. We then calculate the vector of (binomially-distributed) probabilities for each possible group, composed of $k$ cooperators and $N - k$ defectors. We then compute the vector of conditional probabilities ($k/N$) of being a cooperator in each combination. We compute the average payoff to the cooperators by weighing the payoff for cooperation by the probability of being a cooperator, as described. And, {\it mutatis mutandis} for the average payoff to defectors. The mean payoffs capture the expectation if the entire population, $Z$, were randomly paired into groups of size $N$. See Appendix~\ref{App:B} for formal details.

\subsection*{Dynamics.}

We consider the dynamics of small, finite populations. The dynamics determine how the population changes based on assumptions about how the payoffs affect the {\it fitness} of strategies, given what others are doing. Our model uses a dynamics called the {\it Fermi process} \citep{Traulsen-Hauert-2009}. This involves a pairwise comparison where two individuals---a {\it focal individual} and a {\it role model}---are sampled from the population at random. The focal individual copies the strategy of the role model with probability $p$, depending on a comparison of the payoffs of those strategies. If both individuals have the same payoff, the focal individual randomises between the two strategies.\footnote{Note, then, that the focal individual does not always switch to a better strategy; the individual may switch to one that is strictly worse.} 
The probability is a nonlinear function of the payoff difference for $p$, called the {\it Fermi function} \citep{Dirac-1926, Fermi-1926},
\begin{equation}\label{eq:FermiFunc}
    p = \left[ 1 + e^{\lambda (\pi_{f} - \pi_{r})} \right]^{-1},
\end{equation}
where $\lambda \in \mathbb{R}^{+}$ is the intensity of selection, which specifies the importance of neutral drift compared to the selection dynamics.
\footnote{Note that when $\lambda=0$, selection is random; when selection is {\it weak} ($w \ll 1$), $p$ reduces to a linear function of the payoff difference; when $\lambda = 1$, our model gives us back the {\it replicator dynamic}; and, when $\lambda \rightarrow \infty$, we get the {\it best-response} dynamic \citep{Fudenberg-Tirole-1991}.} %
$\pi_{f}, \pi_{r}$ are the payoffs of the focal individual and the role model, respectively. In addition to the parameters described above, we further assume some small rate of mutation, $\mu$, which can be interpreted as mutation, experimentation, error, or noise. For all the results we present in Section~\ref{sec:Results}, we set $\lambda=5, \mu = 0.10$.

In our model, we specify the transition probabilities of the population changing from a state containing $n_C^{Z}$ cooperators to one with $(n_{C}^{Z} + 1)$ or $(n_{C}^{Z} - 1)$ cooperators (i.e., the transition from the dynamics yielding one more or one fewer cooperator in the population at large). A transition matrix (for all possible states) is populated by recursive application of the transition probabilities. The transition matrix is used to compute the gradient of selection for our dynamics. This is defined by the difference between the probability that the number of cooperators increases and the probability that the number of cooperators decreases. We compute the stationary distribution of the process using standard techniques \citep{Sandholm-2007}.\footnote{Note that we already know that the stationary distribution {\it exists}, because the addition of mutation makes the Markov process ergodic. That is to say, the Fermi process was already finite and aperiodic; with mutation, it is also irreducible (i.e., has only one recursive class). This is because there is a positive probability path between any two states, and in the limit every state will be visited an infinite number of times. In the absence of mutation, there are two (singleton) recursive classes corresponding to the two absorbing states where the population is composed entirely of defectors or entirely of cooperators. From being ergodic, it follows that the limit distribution is independent of any initial distributions.} 
Again, see Appendix~\ref{App:B} for formal details.


\section{Results}
\label{sec:Results}

In this section, we examine the results of our model under simulation.\footnote{Our code outputs visual graphics of the selection gradient, average payoffs to each strategy, and the stationary distribution. 
All of our simulation code is available online at \href{https://amohseni.shinyapps.io/tragedies-of-the-commons/}{https://amohseni.shinyapps.io/tragedies-of-the-commons/}.} 
This is useful since human behaviour involves so many degrees of freedom that meaningful analytic results are often unlikely to be obtained. As is common practice in game-theoretic models of social behaviour, we are primarily interested in high-level qualitative results---such as whether cooperation will tend to succeed---rather than exact quantities. 

\subsection*{Simulation Parameter Values.} In our simulations, we fix the population, $Z=100$, and the endowment, $b=1$. All other parameters are variable: this includes the size of the groups, $N\in \{ 1, 2, \ldots, 20 \}$; the perceived risk of negative consequences occurring if cooperation fails, $r \in [0,1]$; the cost of those consequences, understood in terms of the proportion of endowments lost, if cooperation fails, $m\in [0,1]$; the proportion of the endowment that is contributed by cooperators, $c\in [0,1]$; and the critical threshold of cooperators required in order for cooperation to succeed, $p^{*} \in [0,1]$, which determines the number of cooperators required, $n^{*} = \lceil p^{*} \cdot N   \rceil$. 

\subsection*{Qualitative Dynamics.} 
\begin{figure}[htb!]
    \centering
    \begin{subfigure}[t]{0.45\columnwidth}
        \centering
        \includegraphics[width=\linewidth]{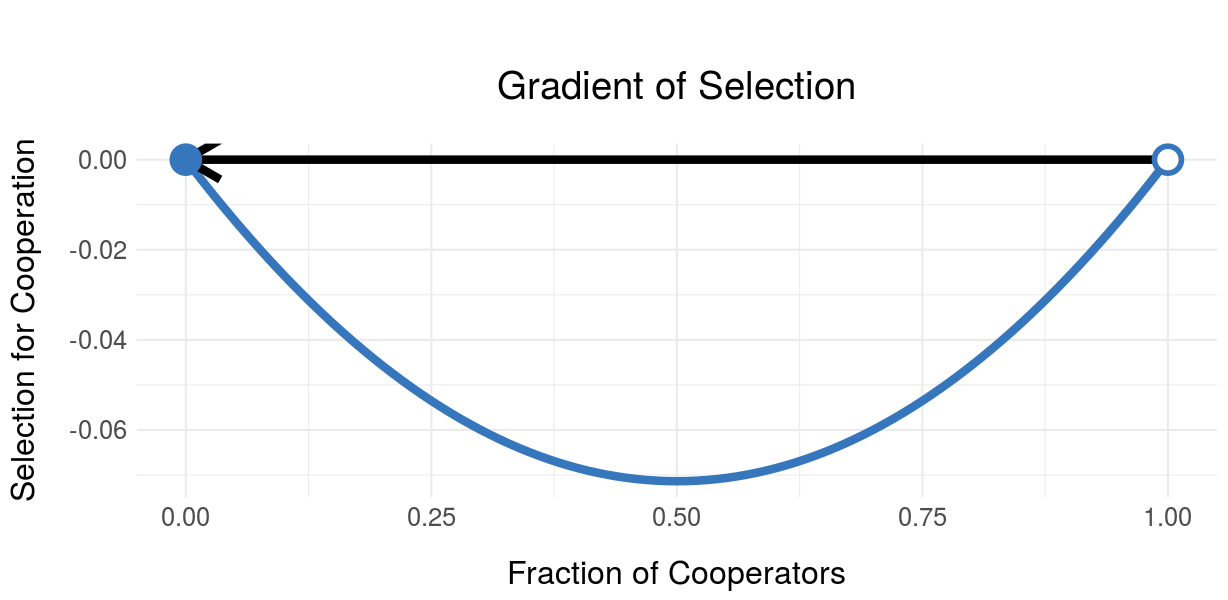}
    \end{subfigure}%
    ~ 
    \begin{subfigure}[t]{0.45\columnwidth}
        \centering
        \includegraphics[width=\linewidth]{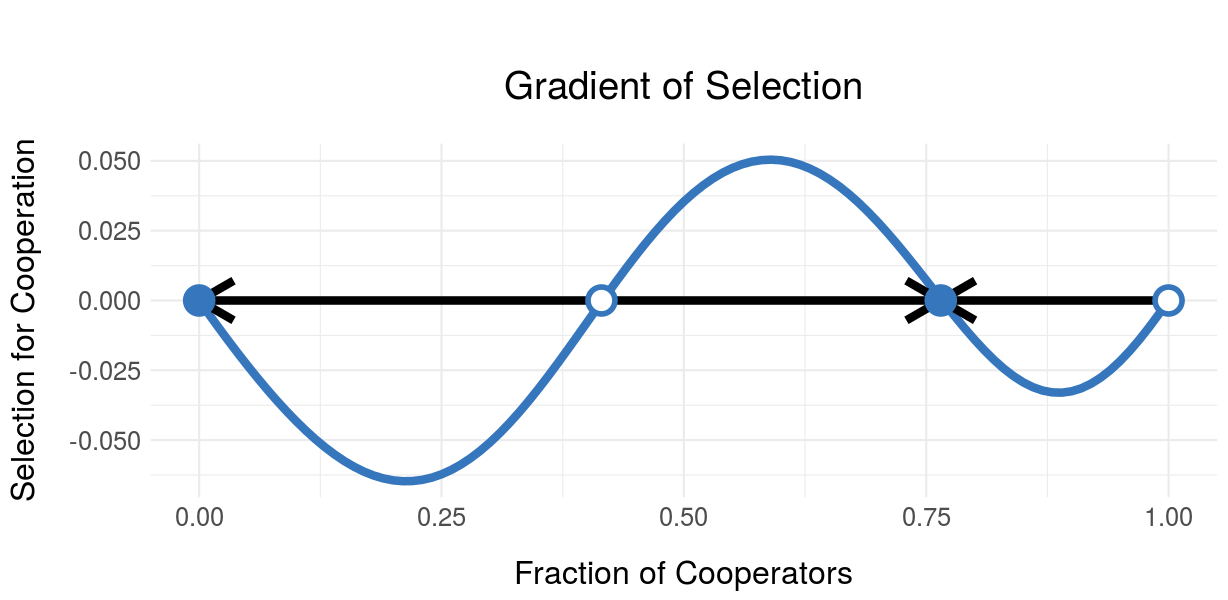}
    \end{subfigure}
    
    \begin{subfigure}[b]{0.45\columnwidth}
        \centering
        \includegraphics[width=\linewidth]{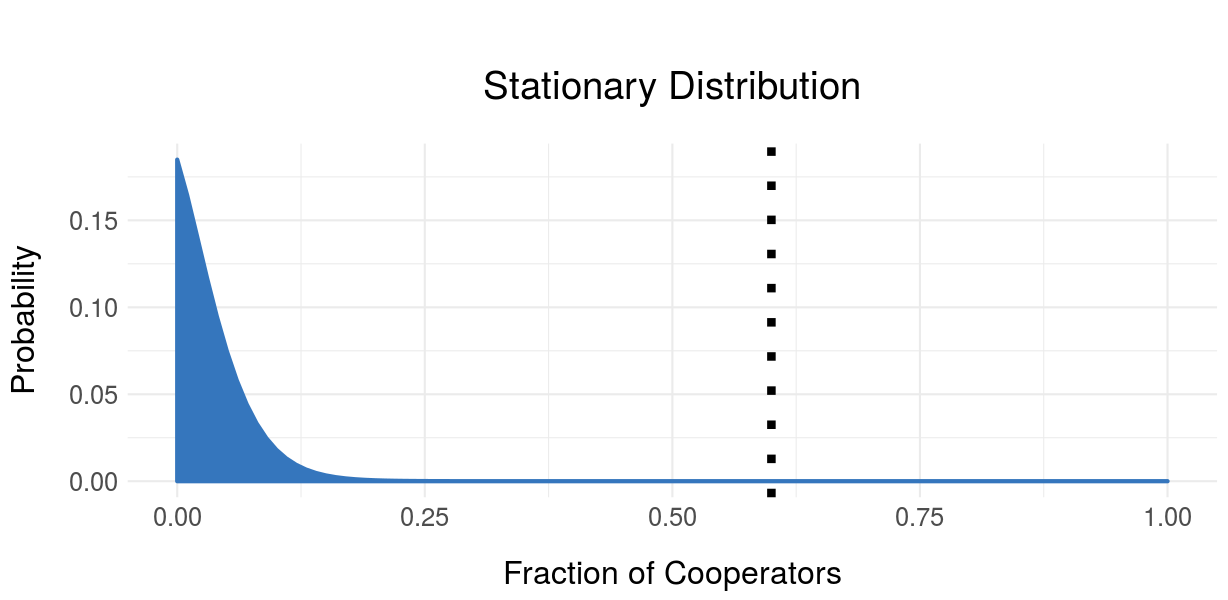}
        \caption{}
        \label{fig:1abot}
    \end{subfigure}%
    ~ 
    \begin{subfigure}[b]{0.45\columnwidth}
        \centering
        \includegraphics[width=\linewidth]{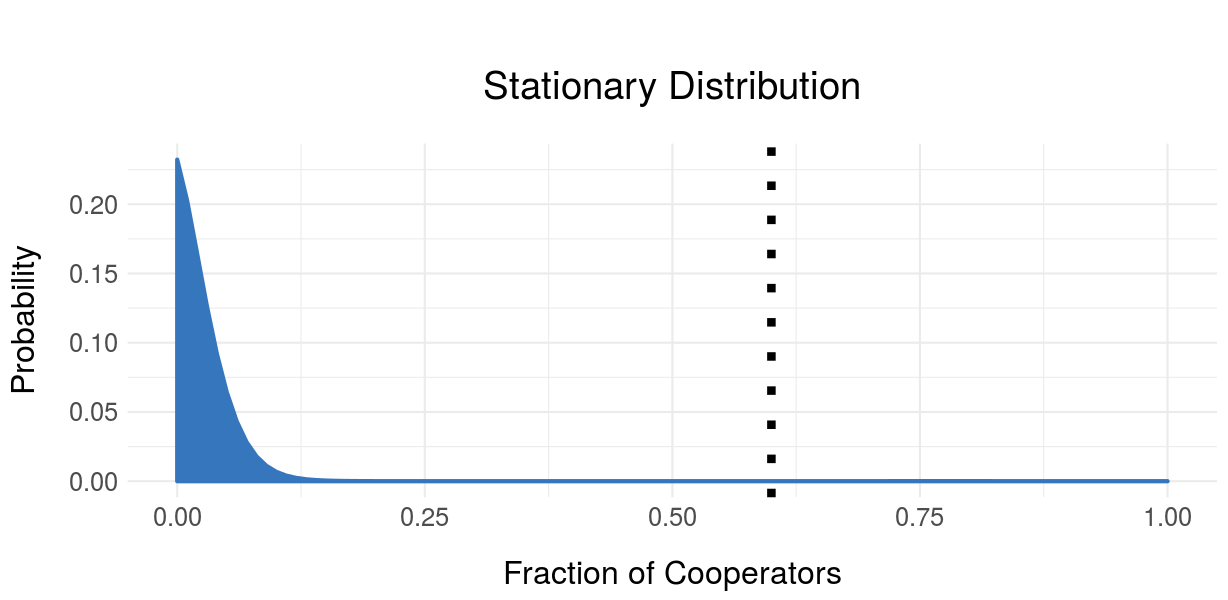}
        \caption{}
        \label{fig:1bbot}
    \end{subfigure}
    
    \begin{subfigure}[t]{0.45\columnwidth}
        \centering
        \includegraphics[width=\linewidth]{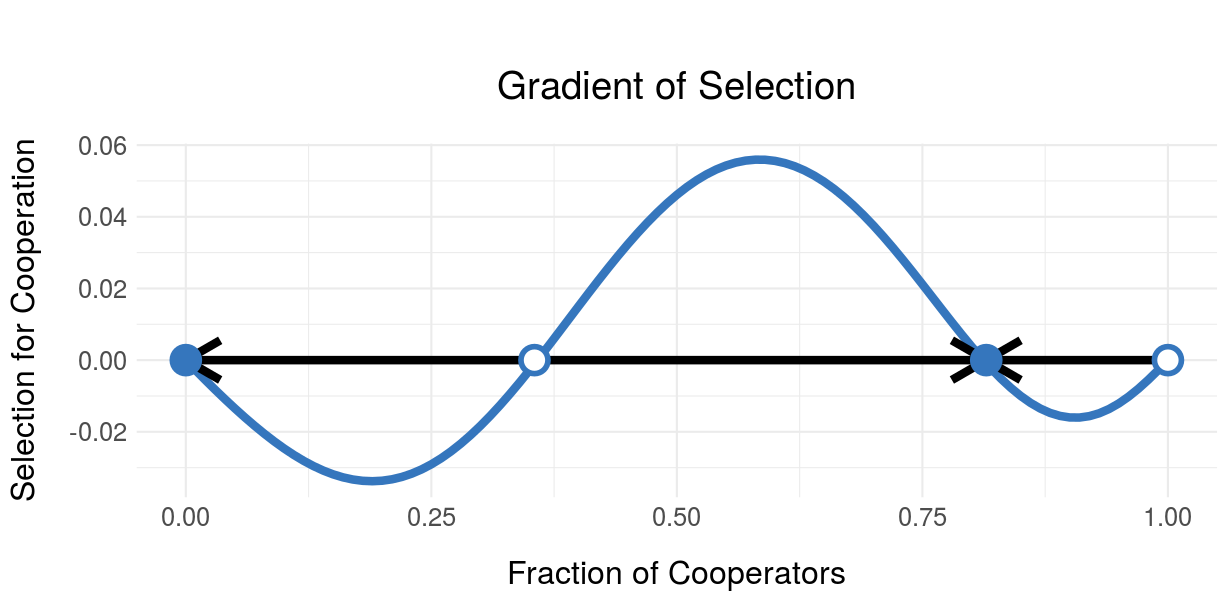}
    \end{subfigure}
    ~ 
    \begin{subfigure}[t]{0.45\columnwidth}
        \centering
        \includegraphics[width=\linewidth]{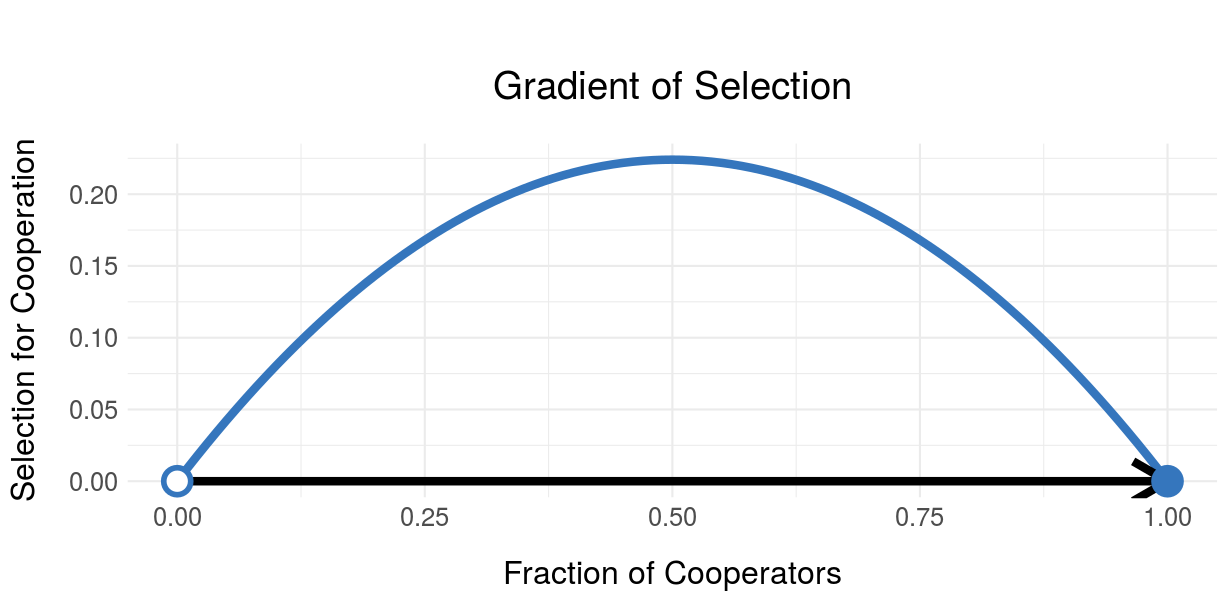}
    \end{subfigure}

    \begin{subfigure}[b]{0.45\columnwidth}
        \centering
        \includegraphics[width=\linewidth]{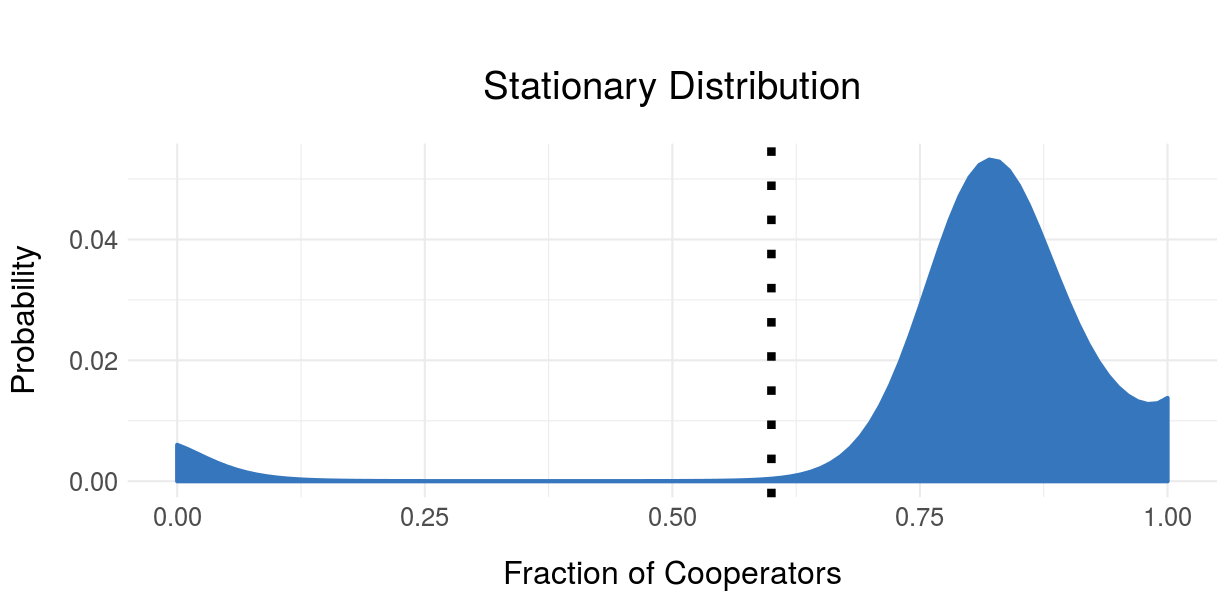}
        \caption{}
        \label{fig:1cbot}
    \end{subfigure}
    ~ 
    \begin{subfigure}[b]{0.45\columnwidth}
        \centering
        \includegraphics[width=\linewidth]{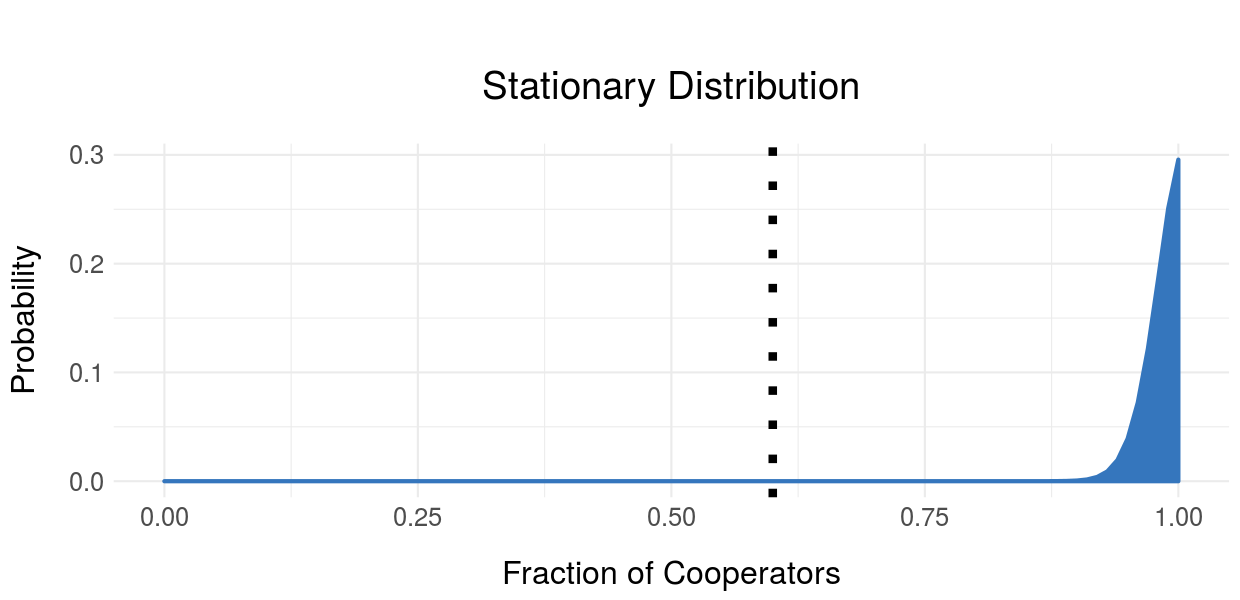}
        \caption{}
        \label{fig:1dbot}
    \end{subfigure}
    \caption{Examples of each qualitative dynamic for the gradient of selection (top) and stationary distribution (bottom) of the mean-field dynamics, resulting in (a) a prisoner's dilemma favouring {\sc all defect}; 
    (b) a bi-stable dynamics favouring the {\sc all defect} equilibrium; 
    (c) a bi-stable dynamics favouring the {\sc polymorphic} equilibrium; 
    and (d) a prisoner's delight favouring {\sc all cooperate}.
    }
    \label{fig:Equilibria-Examples}
\end{figure}

We compute the mean payoff to each strategy, the gradient of selection, and the stationary distribution for the dynamics as described in Section~\ref{sec:Model}. The gradient of selection gives the probable direction of evolution in the short run, whereas the stationary distribution captures the long-run proportion of time that the process spends at any given state. This is the appropriate method of analysis for such finite-population dynamics \citep{Mohseni-2019, Traulsen-Hauert-2009}; see Figure~\ref{fig:Equilibria-Examples} for examples.


The gradient may change from selection for an equilibrium where {\sc all defect} is the unique stable state (top of \ref{fig:1abot}), displaying the qualitative dynamics of a {\it prisoner's dilemma} \citep{Serrano-Feldman-2013}; to a bi-stable dynamics with both an {\sc all defect} equilibrium and a {\sc polymorphic} equilibrium, where both strategies can coexist in the population (top of \ref{fig:1bbot}, \ref{fig:1cbot}---the latter corresponds to an {\it anti-coordination} game \citep{Rapoport-Chammah-1966}); to (in rare circumstances) a unique equilibrium of {\sc all cooperate} (top of \ref{fig:1dbot}), where the qualitative dynamics yield a {\it prisoner's delight} \citep{Binmore-2004}. 

When there is a unique stable state, the stationary distribution reflects that the process spends the majority of its time near this state (bottom of \ref{fig:1abot},~\ref{fig:1dbot}). However, when there are multiple stable states (\ref{fig:1bbot},~\ref{fig:1cbot}), the dynamics may or may not guarantee population-level success relative to the demands of the cooperative challenge---the stationary distribution shows us the proportion of time spent at each state. Hence, although a mixed-population of cooperators and defectors is {\it stable} in Figure~\ref{fig:1bbot}, the stationary distribution indicates that the stable configuration of {\sc all defect} is most probable.

In sum, there are four possible qualitative outcomes for our population under the dynamics described: (1) {\sc all defect} is the unique stable state, near which the process spends all its time; (2) both {\sc all defect} and a {\sc polymorphic} state in which both cooperators and defectors co-exists are stable, yet the process spends a majority of its time near the {\sc all defect} equilibrium; (3) both {\sc all defect} and a {\sc polymorphic} state are stable, but the process spends the majority of its time near the polymorphic mixture; and (4) {\sc all cooperate} is uniquely stable and the process tends toward this state.


\subsection*{Simulation Results.} We want to know how the parameters of interest interact with the achievement of cooperative success (the situations shown in Figures~\ref{fig:1cbot} or~\ref{fig:1dbot}), and which of those will have the greatest effect. Our dynamics encode a process by which strategies that are more successful proliferate via imitation; so, for cooperation to succeed it must be that cooperators do better on average than defectors. Each of the parameters will either work for or against cooperative success. We will examine their respective effects by first considering their extremal values. 

It should be obvious that when the cost of cooperation is maximal ($c=1$), {\it defect} ($D$) will dominate since cooperators have nothing to gain from cooperation in this scenario. Therefore, the lower the cost of cooperation, the more likely individuals will be to cooperate. However, we will see that even a low cost of cooperation does not ensure cooperative success.

\begin{figure}[htb!]
\centering
    \begin{subfigure}[b]{0.35\textwidth}
        \centering
        \includegraphics[width=\linewidth]{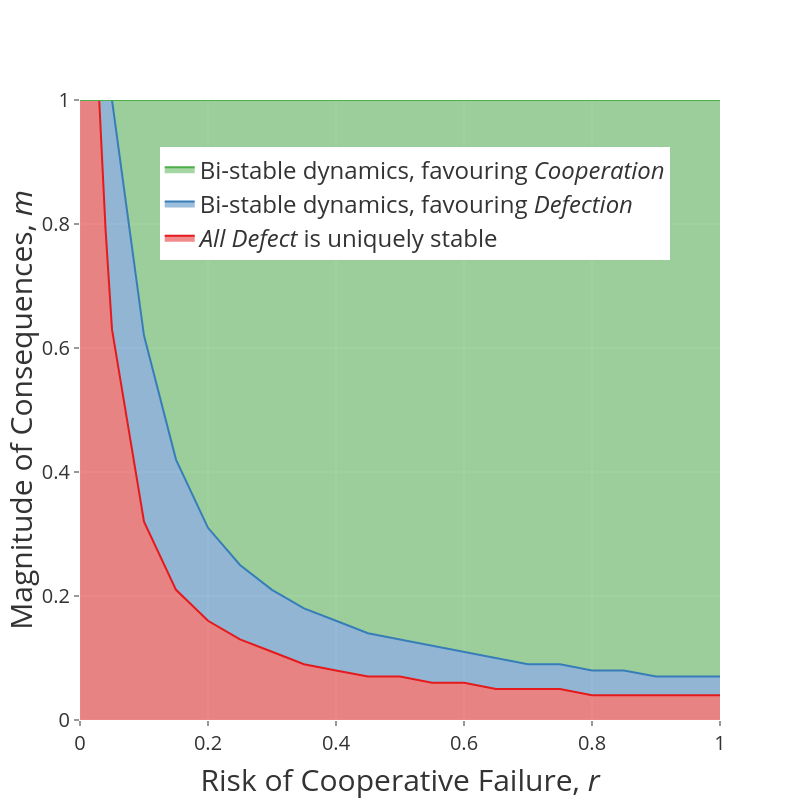}
        \caption{$N=2$}
    \end{subfigure}%
    ~ 
    \begin{subfigure}[b]{0.35\textwidth}
        \centering
        \includegraphics[width=\linewidth]{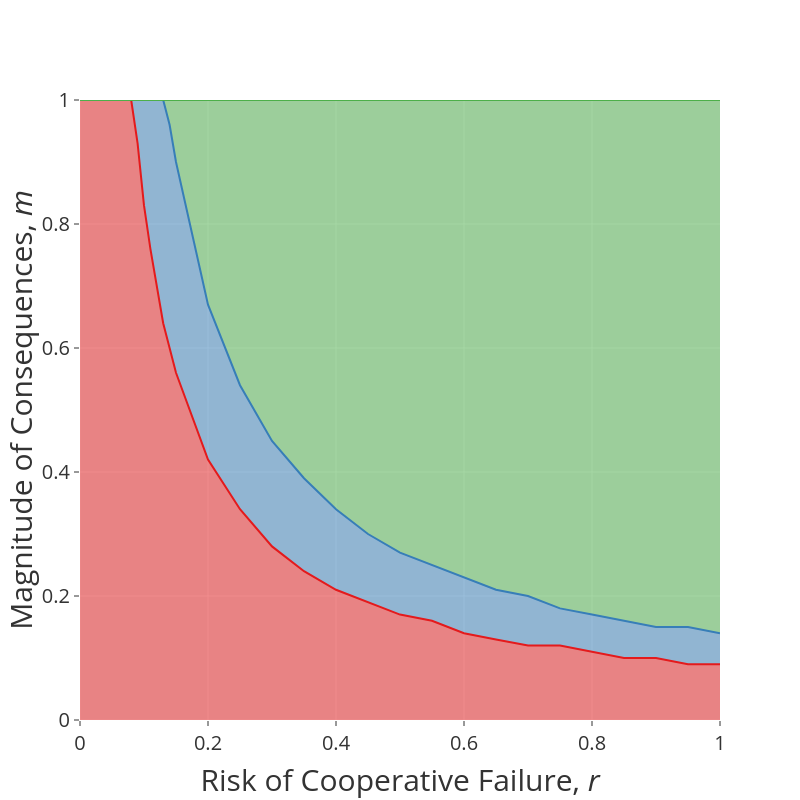}
        \caption{$N=5$}
    \end{subfigure}
    
    \begin{subfigure}[b]{0.35\textwidth}
        \centering
        \includegraphics[width=\linewidth]{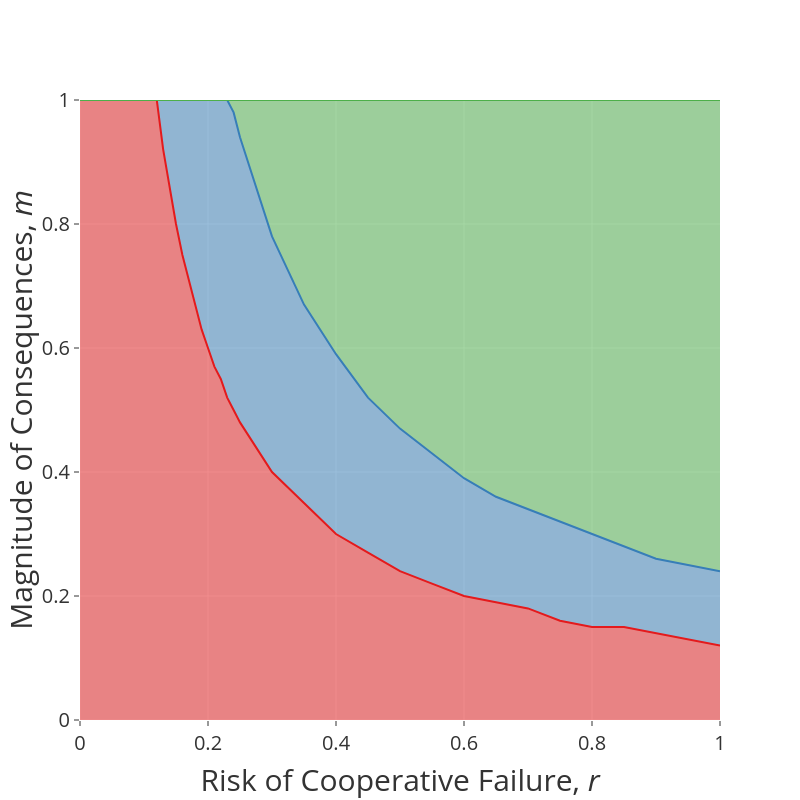}
    \caption{$N=10$}
    \end{subfigure}
    ~
    \begin{subfigure}[b]{0.35\textwidth}
        \centering
        \includegraphics[width=\linewidth]{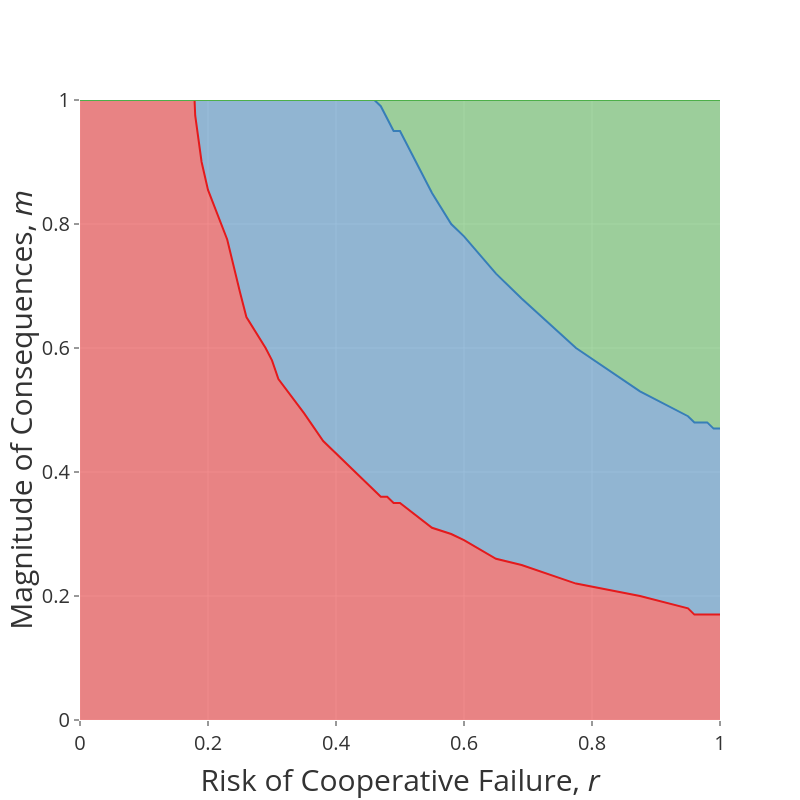}
    \caption{$N=20$}
    \end{subfigure}
    \caption{Qualitative dynamics as a function of the risk, $r$, and magnitude, $m$, of the consequences of a failure to cooperate. 
    In each case, $p^{*}=0.5, c=0.01$.}
    \label{fig:Bifurcation-Examples}
\end{figure}

At the other extreme, when the cost of cooperation is minimal ($c=0$), {\it cooperate} ($C$) is the strictly dominant strategy in almost all cases---the exception is when there is no benefit to cooperation ($r \cdot m=0$), and so the payoffs for $C$ and $D$ are equivalent. On the other hand, if the cost of cooperation is nonzero ($c>0$), and there is no payoff for cooperative success ($r \cdot m=0$), then $D$ is strictly dominant.
%
%
%
That is, improbable but highly consequential failures to cooperate and highly probable but inconsequential failures to cooperate will evoke the same cooperative (or non-cooperative) response. Transitions between qualitative dynamics as a function of the risk, $r$, and magnitude, $m$, of the consequences of cooperative failure are given in Figure~\ref{fig:Bifurcation-Examples}. 

\begin{figure}[htb!]
\centering
    \begin{subfigure}[b]{0.35\textwidth}
        \centering
        \includegraphics[width=\linewidth]{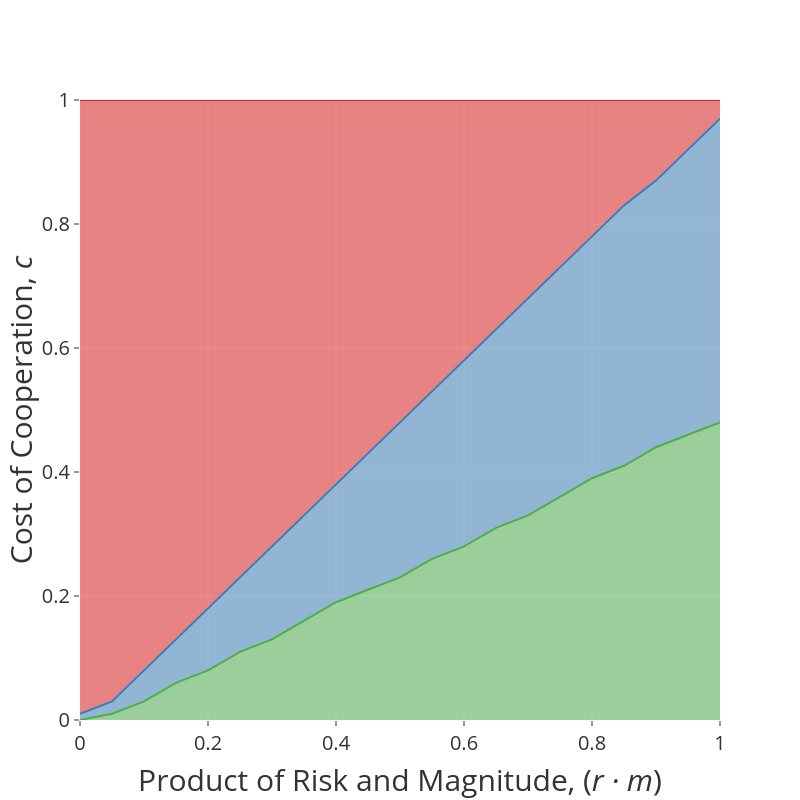}
        \caption{$N=2$}
    \end{subfigure}%
    ~ 
    \begin{subfigure}[b]{0.35\textwidth}
        \centering
        \includegraphics[width=\linewidth]{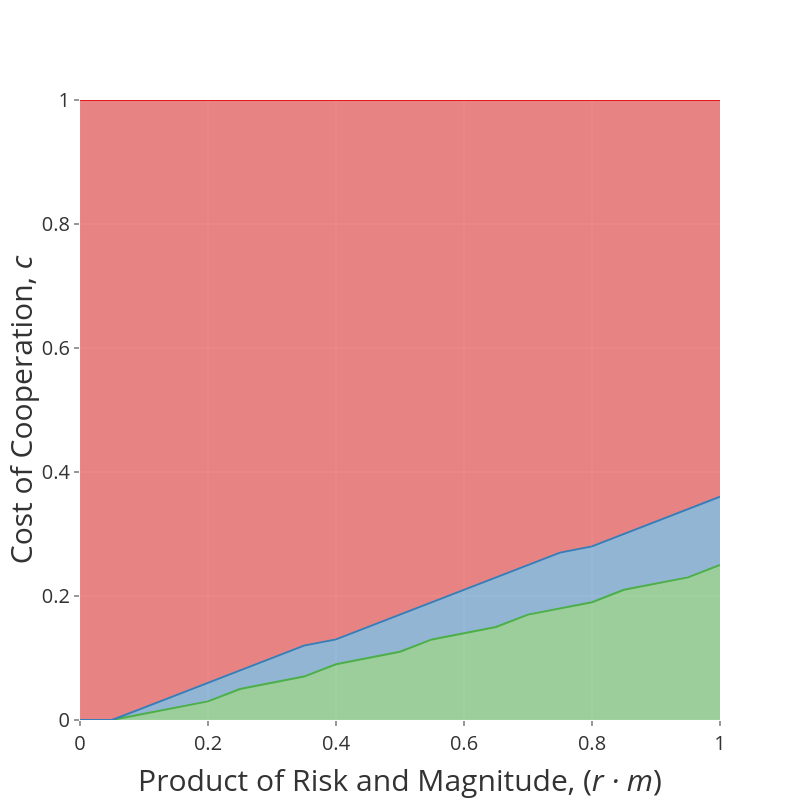}
        \caption{$N=5$}
    \end{subfigure}
    
    \begin{subfigure}[b]{0.35\textwidth}
        \centering
        \includegraphics[width=\linewidth]{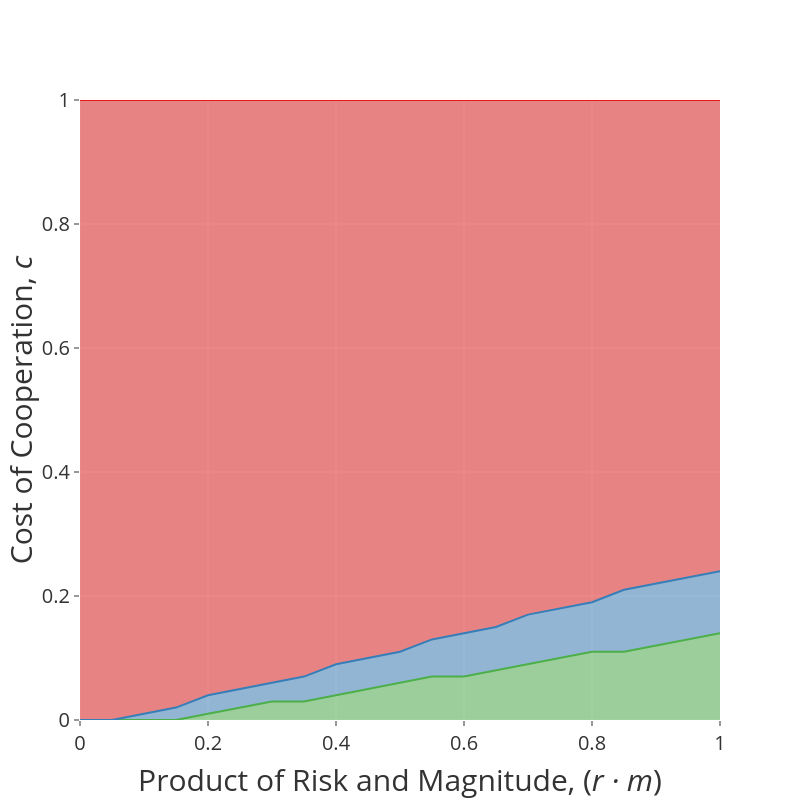}
        \caption{$N=10$}
    \end{subfigure}
    ~
    \begin{subfigure}[b]{0.35\textwidth}
        \centering
        \includegraphics[width=\linewidth]{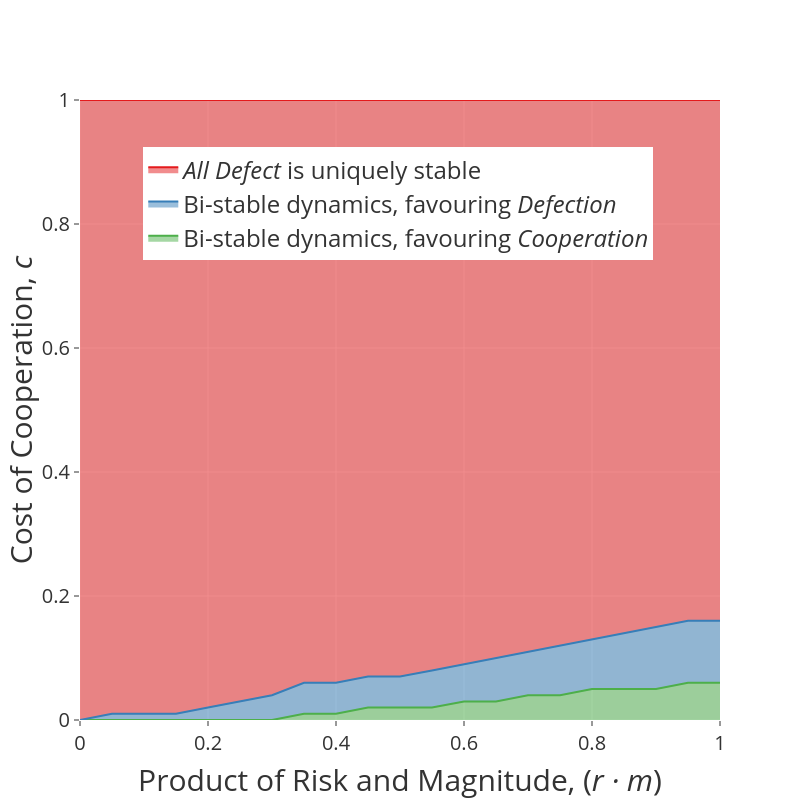}
        \caption{$N=20$}
    \end{subfigure}
    \caption{Qualitative dynamics as a function of the product of the risk, $r$, and magnitude, $m$, of the consequences to fail to cooperate ($r \cdot m$) and the cost of cooperation, $c$. 
    In each case, $p^{*}=0.50$.}
    \label{fig:Bifurcation-RMC}
\end{figure}

Further, Figure~\ref{fig:Bifurcation-RMC} makes clear that cooperative success requires the ratio of the benefit of avoiding the failure to cooperate ($r \cdot m$) and the cost of cooperation ($c$) to be sufficiently favourable.

Figures~\ref{fig:Bifurcation-Examples} and~\ref{fig:Bifurcation-RMC} also illustrate the critical role that group size plays in the prospects of cooperative success.
In particular, all else equal, smaller group sizes are more favourable for cooperation. This is perhaps counter-intuitive; however, the result can be explained as follows. Recall that cooperators must obtain higher average payoffs for cooperation to be favoured. Moreover, agents contribute to the prevalence of cooperation in their respective groups by choosing to cooperate (or not).

When group size is minimal ($N=1$), an agent who chooses to cooperate ensures cooperative success by dint of her action alone. Thus, all groups composed of cooperators succeed, and all groups composed of defectors fail. (This holds for non-extremal values; i.e., $c>0$, $m\cdot r>0$, and $p^*>0$.) This is sufficient to ensure selection for cooperation. 

At the other extreme, when the cooperative group grows arbitrarily large ($N \rightarrow \infty$), an agent's action makes a vanishing contribution to the prospect of cooperative success.\footnote{The reader may identify that this strategic structure is analogous to that of the Paradox of Voting \citep{Condorcet-1793}.} Furthermore, by the law of large numbers \citep{Bernoulli-1713, Poisson-1837}, the proportion of cooperators and defectors in large groups will approach the precise proportion of cooperators and defectors in the whole population. Hence, any group will tend to be equally likely to succeed or fail. Yet, in such groups, defectors will obtain higher payoffs by dint of not having paid the cost to cooperate. This is sufficient to ensure selection for defection. 

In between these extremes, smaller group sizes will tend to favour cooperation and disfavour defection. 
%
%
Thus, we find a special instance of the general pattern that correlation between strategic actions conduces to pro-social behaviour \citep{Hamilton-1964a, Hamilton-1964b, Skyrms-1994}, where here the relevant form of correlation is realised by the contribution of the individual's action to the composition of the group's actions.\footnote{Correlation can be realised variously in a social dilemma---e.g., assortative mating \citep{Eshel-Cavalli-Sforza-1982, Hamilton-1971}, kin selection \citep{Hamilton-1963, Maynard-Smith-1964}, homophily \citep{McPherson-et-al-2001}, and network effects \citep{Broere-et-al-2017}, among others. All of these support cooperation insofar as they make cooperators more likely to interact with one another, and less likely to interact with defectors. Although we lack the space to discuss these here, they constitute an important further dimension of our analysis.}

\begin{figure}[htb!]
\centering
    \begin{subfigure}[b]{0.35\textwidth}
        \centering
        \includegraphics[width=\linewidth]{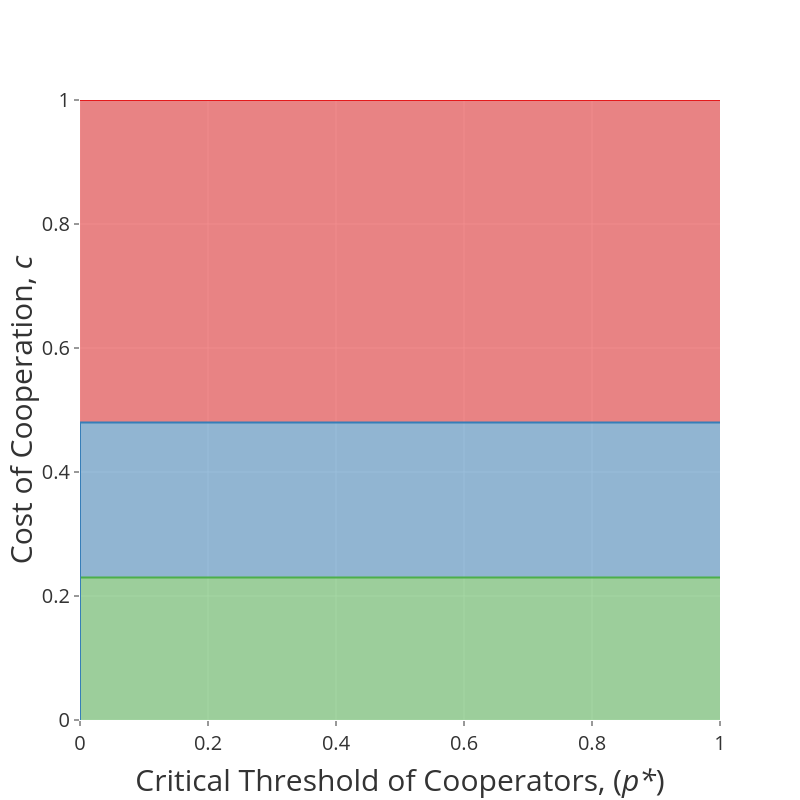}
        \caption{$N=2$}
    \end{subfigure}%
    ~ 
    \begin{subfigure}[b]{0.35\textwidth}
        \centering
        \includegraphics[width=\linewidth]{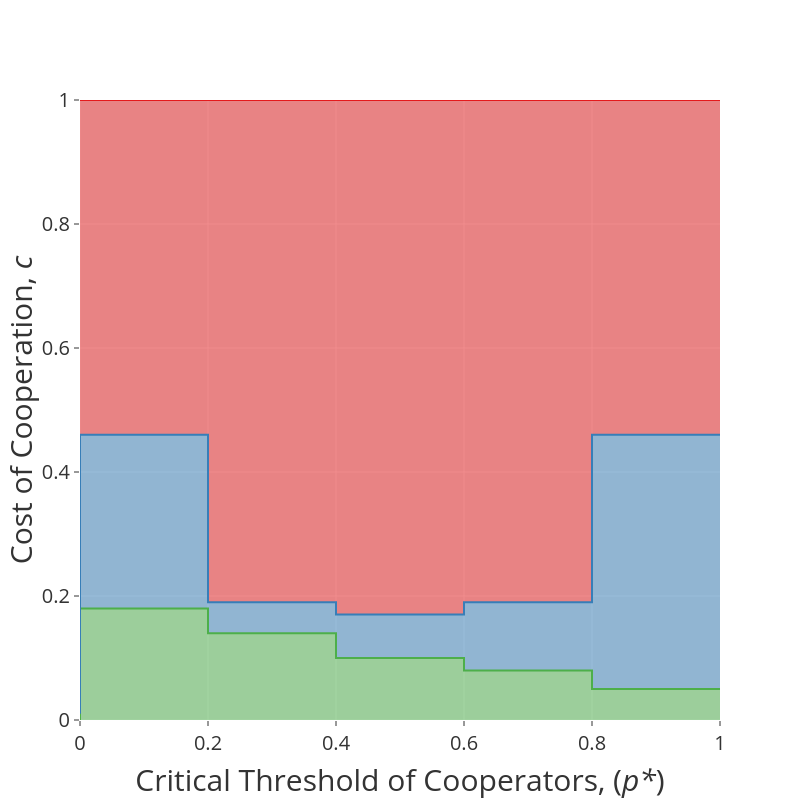}
        \caption{$N=5$}
    \end{subfigure}

    \begin{subfigure}[b]{0.35\textwidth}
        \centering
        \includegraphics[width=\linewidth]{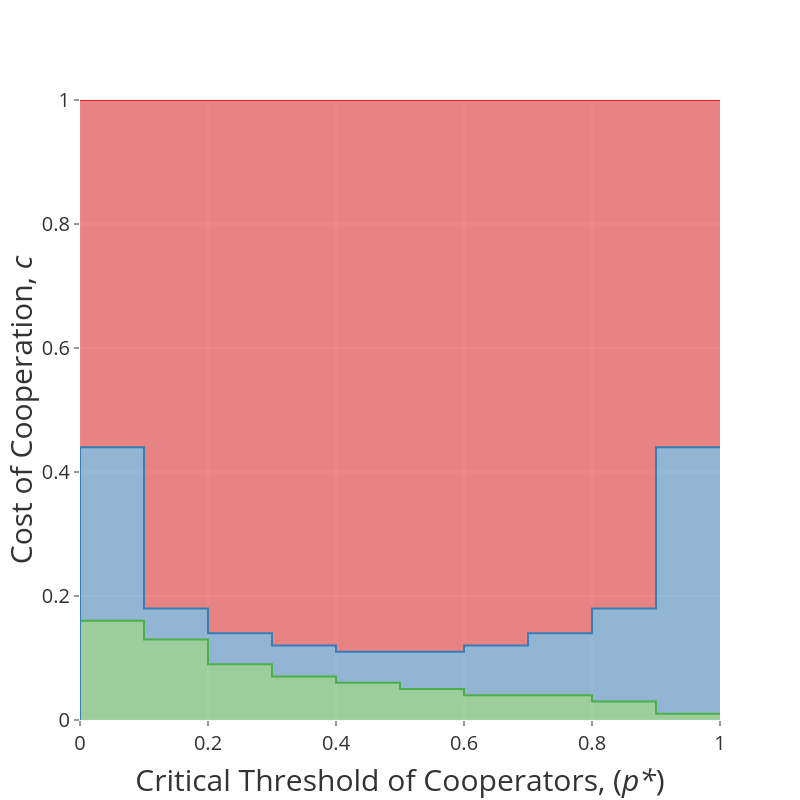}
        \caption{$N=10$}
    \end{subfigure}
    ~
    \begin{subfigure}[b]{0.35\textwidth}
        \centering
        \includegraphics[width=\linewidth]{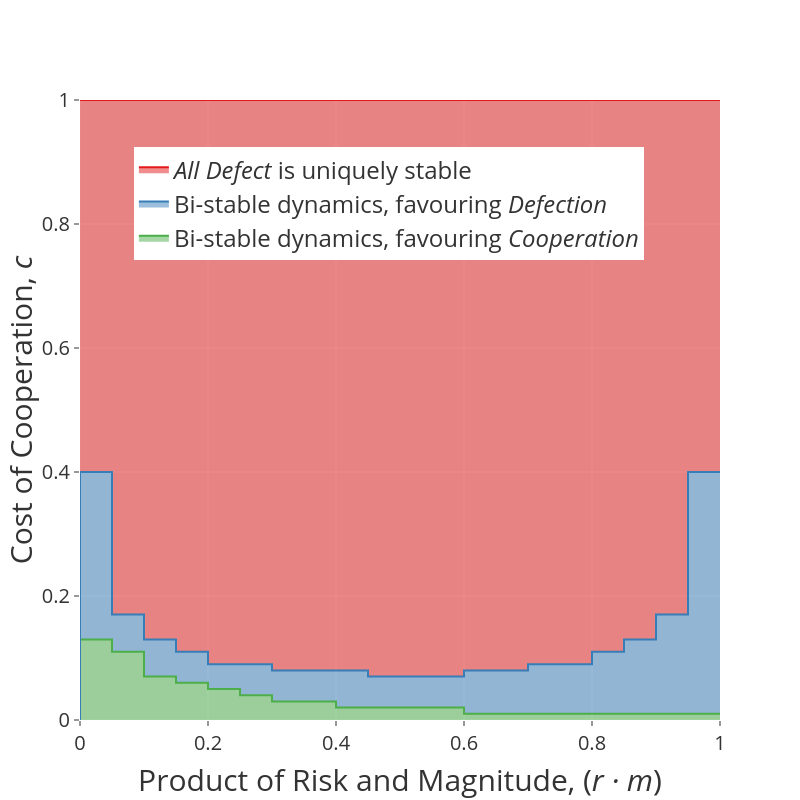}
        \caption{$N=20$}
    \end{subfigure}   
    \caption{Qualitative dynamics as a function of the critical fraction of cooperators required for success, $p^{*}$, and the cost of cooperation, $c$. 
    In each case, $r \cdot m =0.50$.}
    \label{fig:Bifurcation-PC}
\end{figure}

Finally, we can consider the effect of the critical threshold for cooperation, $p^{*}$, on the prospects for cooperative success; see Figure~\ref{fig:Bifurcation-PC}.

Logically, at the extreme where the critical threshold is minimal ($p^{*}=0$), cooperation `succeeds' without need of cooperators, and so defection is dominant. However, when the critical threshold is nonzero ($p^{*}>0$), the general pattern is that the smaller the fraction of cooperators required for cooperation, the more likely cooperation is to succeed. Indeed, sufficiently demanding cooperative endeavours mean that while a {\sc polymorphic} equilibrium in which cooperation co-exists with defection is stable under the gradient of selection, the stationary distribution demonstrates that arriving and remaining at such a cooperative state may be vanishingly unlikely. 

\subsection*{Robustness}


In such models, results need be robust to changes in two sets of parameters: the parameters of the payoff functions, and the parameters of the dynamics. 


The results we describe in Section~\ref{sec:Results} are that the selective advantage of cooperation is monotonically {\it decreasing} in the cost of cooperation, $c$, and the size of cooperative groups, $N$; and, the selective advantage of cooperation is monotonically {\it increasing} in the product of the perceived probability, $r$, and magnitude, $m$, of the consequences of failing to successfully cooperate. 

These findings hold for the entire parameter space of the game. This is demonstrated qualitatively by Figures~\ref{fig:Bifurcation-Examples},~\ref{fig:Bifurcation-RMC}, and~\ref{fig:Bifurcation-PC}; however, these facts can also be proved analytically. See Appendix~\ref{App:Proofs}.

In addition to claims of robustness with respect to the aforementioned parameters, our results are also robust with respect to changes of the parameters of the dynamics---the Fermi process. 


In particular, it is easy to demonstrate that the qualitative dynamics of the game---which determine whether a strategy is being selected for or against---are invariant for the range of conditions characterised by positive, finite intensities of selection, $0 < \lambda < \infty$, and positive mutation rates short of complete randomness, $0 < \mu < 1$.

Thus, the results presented are indeed robust under any reasonable sensitivity analysis.


\section{Conclusion}
\label{sec:Conclusion}

{\it A priori}, one might have thought that inhibitive costs furnish the largest obstacle for cooperative success. Our model provides evidence for this intuition. Thus, our first moral is the obvious one. 

\phantom{a}

\noindent {\bf \textit{Moral 1}}: {\it Lowering the cost of cooperation increases the likelihood of cooperative success.}\footnote{This moral pertains to the likelihood of signing on to an agreement in the first place, but there is also a question of whether individuals who {\it say} they will cooperate in fact do so cooperate \citep{LaCroix-Mohseni-2020}. When signals are cheap, they can be uninformative or dishonest \citep{Crawford-Sobel-1982, Farrell-1987, Farrell-Rabin-1996, Warneryd-1993}. It is well-understood that costly signals can promote honesty \citep{Johnstone-1995, Lachmann-et-al-2001, Pomiankowski-1987, Zahavi-1975, Zahavi-Zahavi-1997}. Further, work in in the {\it economics of identity} suggests that costs for group-membership can inhibit free-riders \citep{Levine-Moreland-1994, Akerlof-Kranton-2000, Akerlof-Kranton-2010, Sosis-Ruffle-2003, Sosis-et-al-2007, Carvalho-2009, Henrich-2009}.}

\phantom{a}

However, our model highlights that the dynamics of cooperation here are more subtle than this. We show that even for small costs, the dynamics may not guarantee population-level success relative to the demands of the cooperative challenge for a range of parameter values.\footnote{In our model, the cost for cooperation is nonnegative. So, we do not account for {\it incentives} to cooperate---i.e., rewards. 
Conversely, we could lower the payoff for defectors by introducing punishment for non-cooperation. This is already something that has been done by, e.g., the ACM \citep{ACM-Enforcement}. Although, empirical data suggests rewards are better than punishments for promoting cooperative behaviour in similar social dilemmas \citep{DeSombre-et-al-2000, Kaniaru-et-al-2007}. Even so, determining what costs/rewards are, how much they are, and how they are distributed is highly nontrivial.} %
Perhaps most surprising is the magnitude of the effect of group size on these outcomes. Thus, we propose the following.

\phantom{a}

\noindent {\bf \textit{Moral 2}}: {\it Small, decentralised groups may benefit sustained cooperation for responsible AI research.}

\phantom{a}

In the context of globalisation, one might have thought that the best approach for fostering AI ethics and safety was to have a universal set of principles to which everyone agrees. However, our model identifies that cooperation spreads as individuals observe other groups succeeding in their cooperative endeavours; this is fostered by the existence of many smaller groups where cooperation can more easily get off the ground. Hence, targeting policy agreements at smaller, independent groups, such as individual research labs or professional organisations, may constitute a more effective path to the aims of the universal adoption of ethical guidelines.

We mention how small group sizes help cooperation \textit{in virtue of} the relationship between positive correlation between strategies and selection for pro-social behaviour. Yet another way to realise such correlation is via free partner choice. When individuals who are inclined to cooperate in adherence to norms are able to freely join one another in their efforts, this can increase their relative likelihood of success and so spread their example throughout the population. Thus, we get the following corollary.


\phantom{a}

\noindent {\bf \textit{Moral 3}}: {\it Voluntary participation in AI policy agreements may catalyse the spread of cooperation.}

\phantom{a}

Individuals choosing to adhere to a set of principles (or not) creates positive correlation. So, affording individuals the option of choosing their cooperative group may be more beneficial than stipulating that one must be a part of the group. However, even in small groups cooperation can fail, as we have demonstrated. Hence,

\phantom{a}

\noindent {\bf \textit{Moral 4}}: {\it It is important to accurately figure both the risks and the consequences of non-cooperation.}

\phantom{a}

The action-relevant parameters in our model are {\it perceived} risk and magnitude. Therefore, it is essential to accurately figure the {\it actual} risks and potential consequences of non-cooperation. Namely, if the perceived risk is significantly lower than the actual risk, then cooperation may fail to obtain despite its (understated) real-world importance. Failing to cooperate, when cooperation matters, incurs a greater loss than cooperating when it was not entirely necessary to do so; hence, there may be less harm in {\it overstating} the negative consequences of failing to cooperate.\footnote{This is a typical line of argument in much of the existential risk literature; see, e.g., \citet{Russell-2019}.} One real-world suggestion that we might glean from this is that it may be beneficial to further incorporate education on AI ethics and safety across a breadth of educational curricula and public outreach efforts.

The insights thus far centre primarily on group dynamics; this says nothing whatsoever about the actual content of ethics guidelines for AI. Of course, the content of these guides can significantly affect the likelihood that community-wide norms are taken up. For example, if the proposed norms are impossible to be instantiated, then they will fail to be adopted. Our final moral addresses this.

\phantom{a}

\noindent {\bf \textit{Moral 5}}: {\it Combining many proposals may undermine their prospects for success.}

\phantom{a}

The likelihood of being willing to adhere to and able to fulfil a set of policies is bounded by the likelihood of adhering to and fulfilling the conjunction of the policies it contains. We show that policies can vary in the magnitude of the cooperative challenge they present. The combination of these facts should make us wary of how demanding we make our policy proposals, if initial and sustained adherence is our aim. Instead, a piecemeal approach in which agents can sign on and gain from participation in single policies may provide a better first step on the path to the ultimate fixation of a robust set of norms for safe and ethical AI.



In closing, we note that even disciplines and fields whose norms of self-regulation are relatively well-entrenched, as in medical practice and biomedical ethics, guiding principles can be ineffective, or at least inefficient, in guiding actual practice when there are competing incentives \citep{Lomas-et-al-1989}.\footnote{Although, this is not to say that the solution is to simply impose {\it hard} laws---this will likely also be ineffective; see discussion in \citet{LaCroix-Bengio-2019}.} This is true even when there is consensus amongst researchers and practitioners about the guidelines themselves \citep{Lomas-1991}---and, it should be apparent that no such consensus exists in AI/ML.

In contrast to other work that has highlighted the inefficacy of non-legislative policy agreements in AI/ML, our analysis focused on the socio-dynamics of cooperation. In doing so, we were able to discuss the circumstances under which we should expect cooperation to arise in the first place. Further, instead of providing normative suggestions for increasing the effectiveness of extant guidelines by varying their content, the normative suggestions that we forward here pertain to the social aspects of cooperation; as such, we do not fall prey to the meta-normative inefficacy described in Section~\ref{sec:Backgrounder}. 

We conclude with a final note about the target of our model, which addresses the proliferation of ethics principles and explores the conditions under which one might expect them to be more or less likely to garner broad adherence or adoption. We do not take our model to diagnose any of these challenges in their entirety; instead, we see it as providing tentative hypotheses that must be combined with empirical evidence and experimentation to inform our judgements adequately.

As a concrete example, the United Nations Educational, Scientific, and Cultural Organisation (UNESCO) voted, in November 2019, to begin a two-year process to create `the first global standard-setting instrument on the ethics of artificial intelligence in the form of a Recommendation’---i.e., a non-legislative policy instrument of the type discussed in Section~\ref{sec:Backgrounder}. An {\it Ad Hoc Expert Group} (AHEG) prepared a first draft of the text for which they solicited public feedback from a wide range of stakeholders over the Summer, 2020. The AHEG will account for this feedback while preparing a final draft, whose adoption will be voted on by UNESCO at the end of 2021. 

We suggest that there may be little reason to think that this document will be much more effective than any of the other 84 documents that have been proposed by NGOs, governments, corporations, professional associations, etc., for the reasons discussed in Section~\ref{sec:Backgrounder}.

So, the target of our paper can be understood, thus. Suppose you are a member of the AHEG, and you are charged with writing a policies guideline for UNESCO. The hope is that UNESCO's 193 member states and 11 associate members will ratify these recommendations (presumably because of wide support from researchers, labs, corporations, etc.  within those member states).  Ratification, then, is a decision to cooperate in the language of our model, and non-ratification is a decision to defect. (Note that ratifying such a document incurs costs to the member state, whereas refusing to do so incurs no such cost.\footnote{Assuming, of course, that the reputational costs incurred by not cooperating are smaller than the costs incurred for cooperating. However, note that reputational costs are endogenous, and are not imposed by the proposal itself.}) 

We can ask how the morals that we have gleaned from our model come to bear on this particular situation in the concrete. According to {\it Moral 1}, it is essential to keep the costs of cooperation low in order to attain the critical mass of cooperators required to make cooperation viable in the first place. Once such a critical mass of cooperators is acquired, the inertia of the system increases the benefits of choosing to cooperate, thus making this option more appealing. 

{\it Moral 2} is inherently difficult in the context of a {\it global} ethics recommendation for AI research, which requires cooperation that spans international borders. We suggested that sustained cooperation is benefited by targeting small, decentralised groups. It should be clear that the member states of UNESCO are neither small nor decentralised. However, it may be possible for UNESCO to provide a {\it framework} of principles for which individual states have significant autonomy to fill in the actual details of how that framework is implemented, or what specific policies are instantiated on the road to achieving the larger goals of the global cooperative agreement. This is analogous to political federalism, like the governmental structure of Canada, wherein there is a division of power between different levels of government---federal, provincial, territorial, or municipal. Increased autonomy for how the principles are instantiated means reduced centralisation and has the added benefit of creating smaller, cooperative subgroups.

{\it Moral 4} suggests that cooperative behaviour may be benefited by soliciting recommendations of both experts and the broader public. It is always the case that there may be some misalignment between those problems that the public deems important versus those that experts in the field do so. When only one of these groups is queried, the relative risks and consequences of failures to cooperate may be undervalued for certain commitments that are deemed important by the other group. As such, even though some risks and consequences may be overstated, this may serve to benefit cooperation in the long run. Furthermore, accurately figuring the risks and consequences of any given scenario will likely require broad interdisciplinary collaboration. So, an AHEG should be composed of experts from a wide range of disciplines (not just computer science), and this expert group may benefit from soliciting public feedback on their proposals. Fortunately, this is precisely the course that UNESCO has taken in this particular instance---the AHEG for the recommendation on the ethics of artificial intelligence is composed of $25$ researchers from as many different countries, and these individuals have diverse backgrounds, including computer and information science, philosophy, law, and engineering \citep{UNESCO-AHEG}. 

Finally, {\it Moral 5} suggests that the proposals should start with modest goals, targeting a small number of specific aspects of safe AI research that ought to be afforded priority. It is always the case that these recommendations can be amended or strengthened in the future. Furthermore, {\it Moral 3} suggests that voluntary participation in such a policy agreement may catalyse the spread of cooperation. Thus, it may be possible to create a positive feedback loop wherein an organisation like UNESCO suggests a small number of modest proposals which incur only a low cost for cooperators to agree. These aspects of the agreement serve to incentivise individuals to voluntarily sign on to such an agreement which, in turn, creates positive correlation that promotes further cooperation. This creates a critical mass of cooperators which then provides further endogenous incentives for individuals to cooperate. 

This concrete example demonstrates how the morals that we derive from our modelling work can be applied, and how these various considerations may have interaction effects that may help nurture population-level success relative to the demands of the cooperative challenge we discuss. Of course, some epistemic humility is a virtue in modelling work of the sort we employ. The results derived from mathematical models of socio-dynamic phenomena are best understood as alerting us to the lower-bound on the complexity of those phenomena, and as providing tentative hypotheses that we must combine with empirical evidence and experimentation to adequately inform our judgements. It would be an error to observe the results of such a model and to become convinced that one fully apprehends the cooperative challenge it depicts. 

It would also be a mistake to interpret our results and arguments as a dismissal of current proposals or guidelines for the safe and ethical development of AI. To the contrary, we believe that such efforts are laudable and necessary; but to give ourselves reasonable odds for success, we must appraise ourselves of an understanding of the basic dynamics of such coordination problems. These mistaken interpretations of our results---with their concomitant negative impacts---must be avoided.

From nuclear proliferation and climate change to ethical AI and AI safety, social dilemmas characterise the gulf between us and the futures for which we hope. The potential positive social impacts of our work are plain. Artificial intelligence promises to fundamentally change nearly every facet of our lives, for better or worse. Hence, effective adherence to ethical norms throughout the process of making numerous inevitable advances in AI will make a difference to the tally of instances in which the process promotes or deranges the prospects of human flourishing. 

Our contribution to this count is simple but, we believe, an important step forward for advancing a productive way of framing this cooperative problem. We deploy theoretical tools from evolutionary game theory to analyse the nature of the social dilemma at play in promoting participation in, and adherence to, the proposed policies and norms. We provide insights that, one the one hand, have not obviously informed extant guidelines and policies, and that, on the other hand, correspond to tractable changes in those proposals which may yield significant impact. If we succeed, stakeholders---research laboratories, university departments, collaborative research groups, and so on---aiming to formulate and coordinate around policy guidelines may have a richer awareness of the challenges involved that may inform the scope, scale, and content of such proposals.

\bibliography{Biblio}
\bibliographystyle{apalikelike}


\appendix

\section{Game Theory}
\label{App:GT}

In this brief appendix, we provide some further game-theoretic background than we had space to discuss in Section~\ref{sec:Model}. For more comprehensive introductions to game theory, see, e.g.,  \citet{Aumann-Hart-1992, Aumann-Hart-1994, Aumann-Hart-2002, Maynard-Smith-1982, Neumann-Morgenstern-1944, Weibull-1997, Young-Zamir-2014}. 

\subsection*{Game-Theoretic Analysis of Cooperation and Conflict.}

Cooperative behaviour persists in human and non-human animal populations alike, but it provides something of an evolutionary puzzle \citep{Axelrod-Hamilton-1981, Darwin-1871, Hauert-et-al-2006, Moyano-Sanchez-2009, Nowak-2012, Nowak-et-al-2004, Taylor-et-al-2004}: How can cooperation be maintained despite incentives for non-cooperative behaviour (i.e., defection)? Evolutionary game theory provides useful tools for analysing the evolution of cooperative behaviour quantitatively in both human and non-human animals.\footnote{See, for example,  \citet{Ashcroft-et-al-2014, Fehl-et-al-2011, Gintis-2000, Grujic-et-al-2015, Grujic-et-al-2012, Hofbauer-Sigmund-1998, Hofbauer-Sigmund-2003, Imhof-Nowak-2006, Kurokawa-Ihara-2009, Maynard-Smith-1982, Nowak-Sigmund-2004, Ohtsuki-Nowak-2006, Ohtsuki-Nowak-2008, Rand-Nowak-2013, Traulsen-et-al-2009, Weibull-1997}.} 

Game theory can be used to study the ways in which independent choices between actors interact to produce outcomes.\footnote{See \citet{Ross-SEP-game-theory} for a philosophical overview.} In game theory, a game is determined by the {\it payoffs}. For example, the payoff matrix for a generic, $2\times2$, symmetric, normal form game is displayed in Figure~\ref{fig:Payoffs}.

\begin{figure}[htb!]
    \centering
    \setlength{\extrarowheight}{2pt}
    \begin{tabular}{cc|c|c|}
        & \multicolumn{1}{c}{} & \multicolumn{2}{c}{Player $2$}\\
        & \multicolumn{1}{c}{} & \multicolumn{1}{c}{$C$}  & \multicolumn{1}{c}{$D$} \\\cline{3-4}
        \multirow{2}*{Player $1$}  & $C$ & $(a,a)$ & $(b,c)$ \\\cline{3-4} 
        & $D$ & $(c,b)$ & $(d,d)$ \\\cline{3-4} \multicolumn{4}{c}{}\\
    \end{tabular}
    \caption{Payoff matrix for a generic, $2\times2$, symmetric, normal form game.}
    \label{fig:Payoffs}
\end{figure}

Each actor (Player $1$ and Player $2$) in this example can choose one of two strategies, $C$ or $D$. The payoffs to each of the players are given by the respective entries in each cell---i.e., the first number in the top-right cell ($b$) is the payoff afforded to Player $1$ when she plays $C$ and her partner plays $D$; the second number ($c$) is the payoff afforded to Player $2$ in the same situation (i.e., when Player $2$ plays $D$ and Player $1$ plays $C$). 

As discussed in the paper, social dilemmas are games where ($i$) the payoff to each individual for non-cooperative behaviour is higher than the payoff for cooperative behaviour, and ($ii$) every individual receives a lower payoff when everyone defects than they would have, had everyone cooperated \citep{Dawes-1980}. 

When $c > a > d > b$, in Figure~\ref{fig:Payoffs}, we have a {\it Prisoner's Dilemma}.\footnote{Named and formalised by Canadian mathematician Albert W. Tucker in 1952, based on  Merrill M. Flood and Melvin Dresher's 1950 model; see \citet{Serrano-Feldman-2013}.} Note that when both actors cooperate (i.e., both play $C$), their payoff is higher than if they both defect ($a > d$), thus satisfying criterion ($ii$) mentioned above. However, each actor has an individual incentive to defect (i.e., play $D$) {\it regardless} of what the other actor does; Player $1$ would prefer to defect when Player $2$ cooperates ($c > a$), and she would prefer to defect when Player $2$ defects ($d > b$)---and {\it mutatis mutandis} for Player $2$. This satisfies criterion ($i$) above. 

In this case, we say that {\it defect} is a {\it strictly dominant} strategy for each player, which leads to the unique {\it Nash equilibrium}: $\langle D , D \rangle$---that is, a combination of strategies where no actor can increase her payoff by unilateral deviation from her strategy. The `dilemma' is that mutual cooperation yields a better outcome for all parties than mutual defection, but, from an individual perspective, it is never rational to cooperate.

\subsection*{Evolutionary Game Dynamics.}

In an evolutionary context, the payoffs are identified with {\it reproductive fitness}, so that more-successful strategies are more likely to propagate, reproduce, be replicated, be imitated, etc. This provides a natural way to incorporate {\it dynamics} to the underlying game.

There are two natural interpretations of evolutionary game dynamics. The first is {\it biological}, where strategies are encoded in the genome of individuals, and those who are successful pass on their genes at higher rates; the second is {\it cultural}, where successful behaviours are reproduced through learning and imitation. We are primarily concerned with processes of cultural evolution. This process should be familiar to those in AI/ML who work on multi-agent reinforcement learning (MARL) \citep{Littman-1994, Shapley-1953, Zhang-et-al-2019}.

In addition to the game, an evolutionary model requires a specification of the {\it dynamics}---namely, a set of rules for determining how the strategies of actors in a population will update (under a cultural interpretation), or how the proportions of strategies being played in the population will shift as they proliferate or are driven to extinction (under a biological interpretation). Evolutionary game dynamics are often studied in infinite populations using deterministic differential equations. For example, the {\it replicator dynamic} \citep{Taylor-Jonker-1978} captures how strategies with higher-than-average fitness tend to increase, and strategies with lower-than-average fitness tend to decrease. A population state is {\it evolutionarily stable} only if it is an asymptotically stable rest point of the dynamics \citep{Maynard-Smith-1982}. 

\subsection*{Stochastic Game Dynamics.}

In finite populations, {\it stochastic} game dynamics are used to study the selection of traits with frequency-dependent fitness \citep{Liu-et-al-2011, Ohtsuki-et-al-2007, Sigmund-2010, Szabo-et-al-2009, Szolnoki-et-al-2009, Traulsen-Nowak-Pacheco-2006}. 

A standard stochastic game dynamics that is used extensively is the {\it Moran process}. This is a simple birth-death process where an individual is chosen proportional to their fitness and replaces a randomly chosen individual with an offspring of its own type \citep{Altrock-Traulsen-2009, Claussen-Traulsen-2005, Huang-Traulsen-2010, Liu-et-al-2017, Liu-et-al-2015, Taylor-et-al-2006, Traulsen-et-al-2007, Wu-et-al-2010, Wu-et-al-2015}.

In the standard {\it Fermi process}, which we discuss in Section~\ref{sec:Model}, an individual is chosen randomly from a finite population, and its reproductive success is evaluated by comparing its payoff to a second, randomly-selected individual from the population \citep{Liu-et-al-2017, Traulsen-Pacheco-Imhof-2006, Traulsen-et-al-2007}.

As mentioned in Section~\ref{sec:Model}, the pairwise comparison of payoffs of the focal individual and the role model informs the probability, $p$, that the focal individual copies the strategy of the role model; the probability function, called the {\it Fermi} function, was presented in Equation~\ref{eq:FermiFunc}, and repeated here for convenience:

$$p = \left[ 1 + e^{\lambda (\pi_{f} - \pi_{r})} \right]^{-1}.$$

Again, if both individuals have the same payoff, the focal individual randomises between the two strategies. Note, then, that the focal individual does not always switch to a better strategy; the individual may switch to one that is strictly worse.

When the intensity of selection $\lambda=0$, selection is random; when selection is {\it weak} ($\lambda \ll 1$), $p$ reduces to a linear function of the payoff difference; when $\lambda = 1$, our model gives us back the {\it replicator dynamic}; and, when $\lambda \rightarrow \infty$, we get the {\it best-response} dynamic \citep{Fudenberg-Tirole-1991}.

Evolutionary game dynamics have been used to shed light upon many aspects of human behaviour, including altruism,\footnote{See, e.g., \citet{Fletcher-Zwick-2007, Gintis-et-al-2003, Sanchez-Cuesta-2005, Trivers-1971}.} moral behaviour,\footnote{See, e.g., \citet{Alexander-2007, Boehm-1982, Harms-Skyrms-2008, Skyrms-2004, Skyrms-1996}.} empathy,\footnote{See, e.g., \citet{Fishman-2006, Page-Nowak-2002}.} social learning,\footnote{See, e.g., \citet{Kameda-Nakanishi-2003, Nakahashi-2007, Rogers-1988, Wakano-Aoki-2006, Wakano-et-al-2004}.} social norms,\footnote{See, e.g., \citet{Axelrod-1986, Bicchieri-2006, Binmore-Samuelson-1994, Chalub-et-al-2006, Kendal-et-al-2006, LaCroix-OConnor-2020, Ostrom-2000}.} and the evolution of communication, proto-language, and compositional syntax,\footnote{See, e.g., \citet{Barrett-2007, Hausken-Hirshleifer-2008, Hurd-1995, Jager-2008, LaCroix-2019-Logic-Game, LaCroix-2020-Dissertation, Nowak-et-al-1999, Pawlowitsch-2007, Pawlowitsch-2008, Skyrms-2010-Signals, Zollman-2005}.}, among many others. See \citet{Ross-SEP-game-theory} for further details.

\section{Technical Details}
\label{App:B}

In this brief appendix, we provide some further formal details for our model than we had space to discuss in Section~\ref{sec:Model}.

\subsection*{Mean Payoffs.}

Recall that the payoffs to each cooperators, $C$, and defectors, $D$, in a group of size $N$ are given as a function of the number of cooperators in that group, $n_C$, as follows:
\begin{align*}
    \pi_{C} (n_C) &= b \cdot \Theta (n_{C} - n^{*}) + b \cdot (1 - rm) \cdot \Theta (n_{C} - n^{*}) - cb, \\
    \pi_{D} (n_C) &= \pi_C (n_C) + cb,
\end{align*}
where $\Theta$ is the Heaviside step function. The mean payoffs to each type in a population of size $Z$, where groups are determined by random mixing, is then given as a function of the total fraction of cooperators in the population, $x_{C}=n_C^Z/Z$, as follows:
{\footnotesize
\begin{align*}
    \Pi_{C} (x_{C}) &= \sum_{n_{C}=0}^{N} \frac{n_{C}}{N} \binom{N}{n_{C}} x_{C}^{n_C} (1 - x_{C})^{N - n_{C}} \pi_{C}(n_C), \\
    \Pi_{D} (x_{C}) &= \sum_{n_C=0}^{N} \frac{N-n_C}{N} \binom{N}{n_{C}} x_{C}^{n_C} (1-x_{C})^{N-n_{C}}\pi_{D}(n_C).
\end{align*}}

\subsection*{Fermi Dynamics.}

The Fermi dynamics uses the average payoffs to each type to determine the probability that a randomly-chosen individual from the population will imitate the strategy of a second randomly-chosen individual from the population. Such a change will produce one of three outcomes: the number of cooperators in the population, $k=n_C^Z$, will increase, decrease, or remain the same. This is captured by the following transition probabilities, which yield a tri-diagonal transition matrix, $T$, for our birth-death process:

\begin{align*}
	T^+(k) &= (1-\mu) \frac{k}{Z}\frac{Z-k}{Z-1}  \left(1 + e^{\lambda (\Pi_{C} - \Pi_{D})} \right) ^{-1} + \frac{\mu}{2}\\
	T^-(k) &=  (1-\mu)  \frac{Z-k}{Z} \frac{k}{Z-1} \left(1 + e^{\lambda (\Pi_{D} - \Pi_{C})} \right) ^{-1} + \frac{\mu}{2} \\
	T^0(k) &=1-T^+(k)-T^-(k)
\end{align*}

where $\lambda$ is the inverse temperature associated with the influence of selection versus drift, and $\mu$ is the rate of mutation. This produces an ergodic Markov process.

\subsection*{Gradient of Selection.}

The gradient of selection of the process captures the expected direction of selection as a function of the number of cooperators in the population, $k$, in a way that is analogous to the mean-field dynamics for the infinite-population case. This is given by

\begin{align*}
    G(k) = T^+(k)-T^-(k) = \frac{k}{Z}\frac{Z-k}{Z-1}\left(\tanh\frac{\lambda}{2}\left(\Pi_C(k)-\Pi_D(k)\right) \right), 
\end{align*}

where $G(k)>0$ implies that selection favours cooperation, and $G(k)<0$ implies that defection is favoured.

\subsection*{Stationary Distribution.}

The stationary distribution of the process captures the long run distribution of time the process spends at each state. For an ergodic process, the stationary distribution is known to be unique and independent of initial conditions of that process. We compute is as follows.

\begin{align*}
    \sigma_k &= \cfrac{ \prod^k_{i=1} \frac{T^+(j-1)}{T^-(j)} }{\sum^{Z}_{i=1} \prod^{i}_{j=1} \frac{T^+(j-1)}{T^-(j)}}\quad \text{for} \ k \in \{ 1, \dots, Z \}.
\end{align*}

The stationary distribution can also be approximated via the Chapman-Kolmogorov equation which states that $n$th step transition matrix corresponds to the $n$th power of  the one-step transition matrix, $T_t=T^t$. Thus, we get that $\sigma$ corresponds to any row of the matrix given by  $\lim_{t\rightarrow \infty} T^t$. 

\section{Proofs}
\label{App:Proofs}

Here we demonstrate several propositions which elucidate the general relationship between selection for cooperation under the Fermi dynamics and the parameters of the strategic interaction.

We say that \textit{selection for} a strategy, $\sigma$, under the dynamics is increasing in parameter $x$ if the transition probability $T^+(k)$ from a state with $k$ individuals playing strategy $\sigma$ to one with $k+1$ individuals playing $\sigma$ increases as $x$ increases, for every interior state. That is $x < x'$ implies $T^+(k;x) < T^+(k;x')$ for all $k \in \{1,\dots,Z-1\}$.

For the following proofs, we fix the initial endowment of agents as some positive constant, $b>0$, without loss of generality, and we assume non-extreme values of the strategic parameters of interest: $N\leq Z \in \mathbb{N}$; $r \in (0,1)$; $m\in (0,1)$; $c\in (0,1)$;  $p^{*} \in (\frac{1}{N},\frac{N-1}{N})$; $\mu \in (0,1)$; and $\lambda \in \mathbb{R}_{>0}$. Note that allowing for extreme values of the parameters makes it so the following inequalities hold only weakly.

    \begin{lemma}\label{lemma}
        Selection for a cooperation under the Fermi dynamics increases (decreases) as the differences of its mean payoff with that of defection increases (decreases).
    \end{lemma}
\begin{quote}
    \begin{proof}
        Recall that the transition probability from a state with $k$ cooperators to one with $k+1$ cooperators is given by $$T^+(k) = (1-\mu) \frac{k}{Z}\frac{Z-k}{Z-1}  \left(1 + e^{\lambda (\Pi_{C} - \Pi_{D})} \right) ^{-1} + \frac{\mu}{2}.$$ So, for any (non-extremal) values of $k$, $\lambda$, and $\mu$, we have it that $T^+$ is proportional to the logit function, $\left(1 + e^{\lambda (\Pi_{C} - \Pi_{D})} \right) ^{-1}$, which in turn clearly increases (decreases) as the difference of mean payoffs, $\Pi_{C} - \Pi_{D}$, increases (decreases).
    \end{proof}
\end{quote}

    \begin{proposition}
        Selection for cooperation decreases as the cost to cooperation increases. 
    \end{proposition}
\begin{quote}
    \begin{proof}
        Consider the difference in mean payoffs between cooperators and defectors, $\Pi_{C}(k;c) - \Pi_{D}(k;c)$, as a function of the cost of cooperation, $c \in (0,1)$, and then fix each of the other parameters at some non-extremal values. 
            
        Observe that for any fixed number of cooperators, $k \in \{1,\dots,Z-1\}$, the difference in mean cost of cooperation $$\Pi_{C}(k;c) - \Pi_{D}(k;c) \propto \sum^N_{n_C=0}\pi_{C}(n_C;c) - \pi_{D}(n_C;c) \propto -cb.$$ Since $b>0$, increasing cost of cooperation, $c$, decreases $\Pi_{C} - \Pi_{D}$, as required. 
            
        By lemma \ref{lemma}, it follows that selection for cooperation decreases as the cost of cooperation increases.
    \end{proof}
\end{quote}

    \begin{proposition}
        Selection for cooperation decreases as the size of cooperative groups increases. 
    \end{proposition}
    \begin{quote}
    \begin{proof}
        Consider the differences in mean payoffs between cooperators and defectors, $\Pi_C(k;N)-\Pi_D(k;N)$, as a function of the size of cooperative groups, $N \in \mathbb{N}$, and fix each other parameter at some non-extremal value.

        Reformulate the difference in mean payoffs in terms of the fraction of the expected fractions of each cooperators and defectors in successful cooperative groups, $p, q$ respectively: $$\Pi_{C} - \Pi_{D} = (p \Pi_{C;s} + (1-p) \Pi_{C;f}) - (q \Pi_{D;s} + (1-q) \Pi_{D;f}),$$ where the subscripts $s$ and $f$ denote when the payoff is for success or failure. We pair the payoffs terms to get $$(p \Pi_{C;s}-(q \Pi_{D;s}) - ((1-p) \Pi_{C;f})-(1-q) \Pi_{D;f}),$$ and then use the fact that $\Pi_{D;s}=\Pi_{C;s}+cb$ and $\Pi_{D;f}=\Pi_{C;f}+cb$, and some algebra, to simplify the expression to: $(p-q)\Pi_{C;s}+(q-p)\Pi_{C;f}-cb$. 

        Since the payoff for success is greater than failure, $\Pi_{C;s}>\Pi_{C;f}$, it follows that the difference in average payoffs is decreasing in the difference of the fraction $p-q$ of successful cooperators and defectors. Taking the derivative of the difference of fractions with respect to $N$  yields  $$\begin{aligned}&\frac{d}{dN}[p(N)-q(N)]\\&=\frac{d}{dN}\left[\sum^{N}_{n_C=\lceil p^*N \rceil}\binom{N}{n_C}(k/Z)^{n_C}(1-k/Z)^{N-n_C} \frac{2n_C-N}{N}\right]\\ & \propto -\sum^{N}_{n_C=\lceil p^*N \rceil}N^{-2} \end{aligned},$$ which is strictly negative. 

        Hence the difference in the fractions of cooperators and defectors who succeed and fail is decreasing in group size, and so the difference in mean payoffs between cooperators and defectors is decreasing. By Lemma \ref{lemma}, it follows that selection for cooperation decreases as group size increases.
    \end{proof}
    \end{quote}

    \begin{proposition}
        Selection for cooperation increases as the product of the perceived risk and magnitude of the consequences of failing to successfully cooperate increases.
    \end{proposition}
\begin{quote}
    \begin{proof}
        Consider the differences in mean payoffs to between cooperators and defectors, $$\Pi_C(k;r,m)-\Pi_D(k;r,m),$$ as a function of the the product of the perceived probability, $0<r<1$, and magnitude, $0<m<1$, such that $0<rm<r'm'<1$.

        Observe that 
        $$\begin{aligned}& \Pi_{C}(k;r,m) - \Pi_{D}(k;r,m) \propto \sum^N_{n_C=0}\pi_{C}(n_C;r,m) - \pi_{D}(n_C;r,m) \\ &=\pi_{C}(0;r,m) - \pi_{D}(0;r,m) + (N-1)cb + \pi_{C}(N;r,m) - \pi_{D}(N;r,m) \\
        &\propto \pi_{D}(N;r,m) - \pi_{C}(0;r,m).\end{aligned}$$ 
        When  $c_N=N$, we have $\pi_{C} (N;r,m) = b(1-c)$ (all cooperators; cooperation succeeds) and when $c_N=0$, we have $\pi_{C}(0;r,m) = b (1 - rm - c)$ (all defectors; cooperation fails). Thus $$\pi_{D}(N;r,m) - \pi_{C}(0;r,m)=brm,$$ and since $b>0$, this is increasing in $rm$.

        Hence the difference in mean payoffs, $\Pi_{C}(k;r,m) - \Pi_{D}(k;r,m)$, is also increasing in $rm$. By lemma \ref{lemma}, it follows that selection for cooperation increases as the product of the risk and magnitude of the failure to cooperate increases.
    \end{proof}
\end{quote}

\end{document}